\pdfoutput=1
\documentclass{JINST}

\let\ifpdf\relax
\usepackage{amsmath}

\newcommand{\Mac}{MAC-E filter}
\newcommand{\celsi}{\ensuremath{^{\circ}\mathrm{C}}}
\newcommand{\ppm}{\mbox{$\mathrm{ppm}/\mathrm{month}$}}

\newcommand{\cps}{\,$\mathrm{counts}/\mathrm{s}$}

\newcommand{\pts}{\mbox{Pt-30-2}}
\newcommand{\lineK}{\mbox{$K$-32}}
\newcommand{\lineL}{\mbox{$L_{1}$-9.4}}
\newcommand{\lineLL}{\mbox{$L_{3}$-32}}

\newcommand{\lineN}{\mbox{$N_{2/3}$-32}}

\newcommand{\kr}{\mbox{$\rm ^{83m}Kr$}}
\newcommand{\krs}{\mbox{$\rm ^{83}Kr$}}

\newcommand{\rb}{\mbox{$\rm ^{83}Rb$}}

\newcommand{\rbkr}{\mbox{$\rm ^{83}Rb/^{83m}Kr$}}
\newcommand{\degree}{\ensuremath{^{\circ}{}}}
\newcommand{\tqdb}[1]{\textquotedblleft #1\textquotedblright}

\title{Ultra-stable implanted $^{\textbf{\small 83}}$Rb/$^{\textbf{\small 83m}}$Kr electron sources for the energy scale monitoring in the KATRIN experiment}

\author{M.~Zbo\v{r}il,$^{a,b,}$\footnote{Corresponding author}\;
S.~Bauer,$^a$
M.~Beck,$^{a,}$\footnote{Present address: Institut f\"ur Physik, Johannes Gutenberg-Universit\"at Mainz, D-55099 Mainz, Germany}\;
J.~Bonn,$^{c,}$\footnote{Deceased}\;
O.~Dragoun,$^b$
J.~Jak\r{u}bek,$^d$
K.~Johnston,$^e$
A.~Koval\'{i}k,$^{b,f}$
E.W.~Otten,$^c$
K.~Schl\"{o}sser,$^g$
M.~Slez\'ak,$^{b,h}$
A.~\v{S}palek,$^b$
T.~Th\"ummler,$^g$
D.~V\'enos,$^b$
J.~\v{Z}emli\v{c}ka$^d$ and C. Weinheimer$^a$\\
\llap{$^a$}Institut f\"ur Kernphysik, Westf\"alische Wilhelms-Universit\"at M\"unster, D-48149 M\"unster, Germany\\
\llap{$^b$}Nuclear Physics Institute of the ASCR,
CZ-25068 \v{R}e\v{z}, Czech Republic\\
\llap{$^c$}Institut f\"ur Physik, Johannes Gutenberg-Universit\"at Mainz,
D-55099 Mainz, Germany\\
\llap{$^d$}Institute of Experimental and Applied Physics, Czech Technical University in Prague,\\
CZ-12800 Prague, Czech Republic\\
\llap{$^e$}Physics Department, CERN,  CH-1211 Geneva 23, Switzerland\\
\llap{$^f$}Laboratory of Nuclear Problems, Joint Institute of Nuclear Research, RU-141980 Dubna, Russia\\
\llap{$^g$}Institut f\"ur Kernphysik, Karlsruher Institut f\"{u}r Technologie,
D-76131 Karlsruhe, Germany\\
\llap{$^h$}Faculty of Mathematics and Physics, Charles University,
CZ-12116 Prague, Czech Republic\\
\\
E-mail: \email{miroslav.zboril@uni-muenster.de}}

\abstract{The KATRIN experiment aims at the direct model-independent determination of the average electron neutrino mass via the measurement of the endpoint region of the tritium beta decay spectrum. The electron spectrometer of the MAC-E filter type is used, requiring very high stability of the electric filtering potential. This work  proves the feasibility of implanted \rbkr\ calibration electron sources which will be utilised in the additional monitor spectrometer sharing the high voltage with the main spectrometer of KATRIN. The source employs conversion electrons of \kr\ which is continuously generated by \rb. The \lineK\ conversion line (kinetic energy of 17.8\,keV, natural line width of 2.7\,eV) is shown to fulfill the KATRIN requirement of the relative energy stability of $\pm$1.6\,ppm$/$month. The sources will serve as a standard tool for continuous monitoring of KATRIN's energy scale stability with sub-ppm precision. They may also be used in other applications where the precise conversion lines can be separated from the low energy spectrum caused by the electron inelastic scattering in the substrate.}


\begin{document}

\section{Introduction}

The KArlsruhe TRItium Neutrino (KATRIN) experiment represents a next-generation tritium $\beta$ decay experiment designed to perform a high precision direct measurement of the electron neutrino mass $m(\nu_\textrm{e})$ with the sensitivity of $0.2\,\mathrm{eV}/c^{2}$ (90\,\% C.L.). In the $\beta$ decay experiment the observable is the weighted squared mass \cite{Ott08}
\begin{equation}
\label{equ:neutrino}
m^2(\nu_\textrm{e}) = \sum\limits_{j=1}^{3}{|U_{\textrm{e} j}|^2 \, m^2(\nu_j)}\,,
\end{equation}
where $U$ denotes the unitary mixing matrix and $\nu_1$, $\nu_2$, $\nu_3$ are the mass eigenstates. The square root of eq.~(\ref{equ:neutrino}) is often called the average electron neutrino mass. KATRIN is a successor experiment of the neutrino mass experiments carried out in Mainz (Germany)~\cite{Pic92-NIM} and Troitsk (Russia)~\cite{Lob85} which set the upper limit on $m(\nu_\mathrm{e})$ of about $2\,\mathrm{eV}/c^2$ \cite{Kra05,Lob03}. Therefore, the aim of the KATRIN experiment \cite{Ott08, KAT04, Dre12} represents the improvement of the neutrino mass sensitivity by one order of magnitude.

For the observation of a non-zero neutrino mass signature in sub-eV range in the endpoint region of tritium $\beta$ spectrum (decay energy $Q=18.6$\,keV) the methods of high resolution electron spectroscopy are necessary together with a high luminosity and very low background level. For this purpose the upcoming KATRIN experiment~\cite{KAT04} uses a windowless gaseous tritium source and two successive electrostatic retardation filters with the magnetic adiabatic collimation (so-called MAC-E filters) \cite{Pic92-NIM,Lob85}. However, the three aforementioned requirements are not the only ones which are connected to the challenging realisation of the  KATRIN experiment. A possible \emph{instability of the energy scale} of the main KATRIN spectrometer is one of the main systematic effects: The principle of the MAC-E filter technique relies on the knowledge of the electric retarding potential which is experienced by the electrons on their path through the spectrometer. Owing to the designed sensitivity, the precise knowledge of the retarding potential at every stage of the measurement is inevitable in the KATRIN experiment. Besides the use of the state-of-the-art equipment for a direct measurement of the high voltage (HV), several very stable calibration electron sources, based on atomic/nuclear standards, will be utilised in KATRIN. One of the electron sources will be continuously measured by an additional MAC-E filter spectrometer (so-called monitor spectrometer which is an upgraded spectrometer of the former Mainz Neutrino Mass Experiment) to which the HV will be applied, corresponding at the same time to the filtering potential of KATRIN. This way a redundant twofold monitoring system will be formed, increasing the confidence in the stability of the main spectrometer energy scale. In this work the feasibility of solid electron source based on the metastable isotope \kr\ is tested. In this type of source the process of internal conversion of \kr\ is utilised, where \kr\ is continuously generated by \rb\ implanted into the substrate. The monitoring task of KATRIN demands the relative energy stability of $\delta E / E \leq \pm1.6$\,ppm$/$month of the \lineK\ conversion line (kinetic energy of $E=17.8$\,keV, natural line width of $\Gamma=2.7$\,eV~\cite{Cam01}).

This article is organised as follows: After briefly introducing the KATRIN experiment, in section~\ref{sec:motivation} we explain the necessity of the energy scale stability and its monitoring. The requirements for an electron source for such monitoring are discussed and it is shown that the implanted \rbkr\ source may fulfill these requirements. In sections~\ref{sec:spectroscopy} and \ref{sec:results} the test measurements carried out at the MAC-E filter spectrometer in the Institut f\"{u}r Physik, Universit\"{a}t Mainz, are described. Firstly the experimental set-up and evaluation methods are introduced (section~\ref{sec:spectroscopy}) and later the experimental results obtained with the first four samples of the implanted \rbkr\ source are summarised (section~\ref{sec:results}). Finally, in section~\ref{sec:discussion} we discuss our results and give an outlook on the further development of the calibration electron sources for KATRIN.

\section{Natural calibration electron standard for KATRIN}
\label{sec:motivation}

\subsection{The MAC-E filter technique}
\label{subsec:mace}

A sensitive search for the neutrino mass by measuring the endpoint region of the tritium $\beta$ decay spectrum requires an electron spectrometer with high luminosity as well as high energy resolution. Both these basic requirements are fulfilled in the concept of magnetic adiabatic collimation with an electrostatic filter, abbreviated as MAC-E filter. This principle, firstly introduced in the 1970s \cite{Hsu76} and early 1980s \cite{Bea80, Bea81, Kru83}, was later adopted for the use in the neutrino mass experiments at Mainz (Germany)~\cite{Pic92-NIM} and Troitsk (Russia)~\cite{Lob85}.

Basically, a spectrometer of the {\Mac} type is a vacuum vessel containing a cascading system of cylindrical HV electrodes and two superconducting solenoids which are placed on both ends of the vessel. The solenoids generate a highly inhomogeneous magnetic field guiding the electrons along the magnetic field lines from their origin in the source to the detector. The magnetic field is symmetrical with respect to the central plane (so-called analysing plane) of the spectrometer. The minimum field strength $B_\mathrm{min}$ at the central plane is reduced by several orders of magnitude with respect to the maximum $B_\mathrm{max}$ occurring at the centre of the solenoids (also called pinch magnets). The magnetic gradient force transforms most of the transverse energy $E_\perp$ of the electron into the longitudinal motion. This is represented as a smooth collimation of the momentum vector with respect to the magnetic field direction. As far as the magnetic field $B$ varies relatively slowly along one cyclotron loop of the electron the energy transformation takes place adiabatically, i.e. the magnetic moment $\mu$ keeps constant\,\footnote{Since the electrons generated in tritium $\beta$ decay reach a maximum value of the relativistic parameter $\gamma = 1.04 \approx 1$, the non-relativistic approximation may be used for this brief explanation of the principle.} \cite{KAT04},
\begin{equation}
 \label{equ:adiabinv-nonrelativistic}
\mu = \frac{E_\perp}{B} = \mathrm{const}\,.
\end{equation}
Thus, the $\beta$ electrons, isotropically emitted from the source of a given area, are transformed into a broad beam of electrons flying almost parallel to magnetic field lines. This parallel beam of electrons is running against an electrostatic potential formed by the set of cylindrical electrodes. All electrons with enough energy to pass the electrostatic barrier are reaccelerated and collimated onto a detector, all others are reflected. Therefore the spectrometer acts as an integrating high-energy pass filter. Varying the electrostatic retarding potential allows the measurement of an electron spectrum in an integrating mode.

The energy resolution of a {\Mac} follows from eq.~(\ref{equ:adiabinv-nonrelativistic}). In the extreme case where the total kinetic energy of the electron at the starting point, $E_\mathrm{start}$, is given in the form of transverse energy, the adiabatic transformation according to eq.~(\ref{equ:adiabinv-nonrelativistic}) will result in a small remainder of transverse energy left over at the analysing plane. This maximum amount of non-analysable energy defines the theoretical resolution $\Delta E$,
\begin{equation}
\label{equ:energy-resolution}
\Delta E = (E_\perp)_\mathrm{max} = E_\mathrm{start} \cdot \frac{B_\mathrm{min}}{B_\mathrm{max}}\,.
\end{equation}
Therefore, the resolution is basically only limited by the minimal ratio $B_\mathrm{min}/B_\mathrm{max}$ of the magnetic fields (at the central plane and at the centre of the solenoids) that can be realised experimentally. In order to suppress electrons which have a very long path within the tritium source and therefore exhibit a high scattering probability, the source is placed in a magnetic field $B_\mathrm{s}<B_\mathrm{max}$. Due to the magnetic mirror effect the maximum accepted starting angle $\theta_\mathrm{start}^\mathrm{max}$ of the electrons is restricted to
\begin{equation}
\label{equ:pinch-angle}
\theta_\mathrm{start}^\mathrm{max} = \arcsin \sqrt{\frac{B_\mathrm{s}}{B_\mathrm{max}}} \,.
\end{equation}

For an isotropically emitting electron source the transmission function of an ideal MAC-E filter can be expressed analytically \cite{Pic92-NIM,KAT04} as a function of the retarding potential $U<0$ and the magnetic field strengths $B_\mathrm{s}$, $B_\mathrm{max}$ and $B_\mathrm{min}$.

\subsection{Motivation for continuous monitoring of the KATRIN energy scale}

The complex set-up of the KATRIN experiment, spanning over 70\,m in length and currently being built and commissioned at the Karlsruher Institut f\"{u}r Technologie (KIT), can be described as follows: The high luminosity windowless gaseous tritium source ensures a constant count rate of $\beta$ electrons which are magnetically guided to the tandem of MAC-E filter spectrometers. The so-called pre-spectrometer rejects the large low energy part of the tritium $\beta$ spectrum which does not carry any significant information for the determination of $m(\nu_\mathrm{e})$. Thus, only the electrons originating from the last $\approx300$\,eV below the tritium $\beta$ spectrum endpoint of $E_0=18.6$\,keV enter the succeeding so-called main spectrometer. Here the high precision energy analysis takes place. The electrons which pass the analysing potential are counted in the detector (silicon PIN diode array of 148 pixels with a total area of 63\,cm$^2$) with the energy resolution of $\simeq1.5$\,keV for 18.6\,keV electrons. The kinetic energy of the $\beta$ electrons will be analysed on the energy scale which is defined by the difference between the tritium source potential and the analysing retarding HV potential of $-18.6$\,kV of the main spectrometer.

The fluctuations of the energy scale represent one of the six main systematic uncertainties of the KATRIN experiment. The KATRIN project requires that none of the individual systematic effects gives the uncertainty contribution of more than about $\Delta m^2_{\mathrm{syst,\;}i} = 0.007\,\mathrm{eV}^{2}/c^4$, thus the total systematic uncertainty (where further minor contributions are included as well) with respect to the observable $m^2(\nu_\mathrm{e})$ is estimated as $\Delta m^2_\mathrm{syst} = 0.017\,\mathrm{eV}^{2}/c^4$ \cite{KAT04}. The source potential will be a few hundreds Volts which can be measured precisely with the help of commercial devices. On the other side, the precise measurement of the HV is generally less reliable as instabilities can occur. The relationship between the instabilities of the retarding HV and the systematic uncertainty of $m^2(\nu_\mathrm{e})$ can be derived by approximating the $\beta$ spectrum by a simplified form in its endpoint vicinity and considering the HV fluctuations, introduced to the measurement. Assuming the HV fluctuations as a Gaussian distribution with a given width $\sigma$ one arrives at \cite{Ott08,Rob88}
\begin{equation}
\label{equ:deltamasse-sigma-zusammenhang}
\Delta m^2_\mathrm{HV}(\nu_\mathrm{e})  c^4 = -2 \sigma^2\,.
\end{equation}
This relationship shows that an \emph{unrecognised} fluctuation of the retarding potential in the form of a Gaussian broadening of the width $\sigma$ imprints on the $\beta$ spectrum in the manner of reducing the neutrino mass squared, i.e. the introduced HV fluctuation affects the $\beta$ spectrum as a shift towards negative $m^2(\nu_\mathrm{e})$ values. Finally, comparing eq.~(\ref{equ:deltamasse-sigma-zusammenhang}) with the KATRIN requirement, the upper limit on the unrecognised fluctuation of the HV scale is obtained:
\begin{equation}
\label{equ:sigma-obergrenze}
\Delta m^2_\mathrm{HV}(\nu_\mathrm{e}) \leq 0.007\,\mathrm{eV}^2/c^{4} \, \Rightarrow \, \sigma \leq 0.059\,\mathrm{eV}\,.
\end{equation}

The upper limit on $\sigma$ stated in eq.~(\ref{equ:sigma-obergrenze}) can be interpreted as the requirement on the HV scale stability of 
\begin{equation}
\Delta(\mathrm{HV})_\mathrm{abs}=\pm60\,\mathrm{mV}
\end{equation}
which represents the relative stability of
\begin{equation}
\label{emu:requirement-relative}
\Delta(\mathrm{HV})_\mathrm{rel}=\pm3.2\,\mathrm{ppm}
\end{equation}
of the retarding potential of 18.6\,kV corresponding to the tritium $\beta$ spectrum endpoint $E_0$. The indicated HV stability must be fulfilled over each KATRIN experimental run spanning about two months\,\footnote{Tritium run times longer than two months will be avoided due to the time needed for the regeneration of special cryogenic components in the experimental set-up.}. A separate value of $m^{2}(\nu_\mathrm{e})$ will be determined from each run and the resulting neutrino mass will be calculated as an average value. Evidently, successful stabilisation of the HV scale over more than one run would allow to directly combine the individual data series and to examine the $\beta$ spectrum at higher statistics. The total live time for achieving the aforementioned neutrino mass sensitivity amounts to three years \cite{KAT04}.

For the purpose of an ultra-precise monitoring of the energy scale stability in KATRIN a twofold monitoring system is currently under construction in the KIT. Firstly, the HV of $-18.6$\,kV applied to the main spectrometer will be measured directly using a specially developed high precision HV divider together with a state-of-the-art digital voltmeter. Secondly, a suitable calibration source of monoenergetic electrons will be continuously measured in the third MAC-E filter spectrometer of KATRIN, the aforementioned monitor spectrometer. This spectrometer will share the HV with the main spectrometer. Any change of the measured energy of the monoenergetic electrons can indicate a problem in the direct measurement of the HV. This way an important redundancy can be achieved by relying not only on high precision devices, but also on the natural standard of atomic and nuclear physics.

The direct measurement of HV with the sub-ppm precision will be accomplished with the help of two ultra-stable custom-made high precision HV dividers \cite{Thu09,dipBauer} developed for KATRIN in the Institut f\"{u}r Kernphysik, Universit\"at M\"unster, in cooperation with the Physikalisch-Technische Bundesanstalt (PTB) Braunschweig. The HV divider scales the HV of $-18.6$\,kV down to low voltage of about $-10$\,V which can be measured by a commercial digital voltmeter with high precision. The voltmeter is regularly calibrated with the help of a commercial 10\,V DC reference which, in turn, is calibrated against PTB's Josephson voltage standard \cite{Koh03}. The HV divider concept is similar to the one used before for producing the PTB's standard divider MT100 \cite{Mar01}, although very different types of resistors were used in these two concepts. Both KATRIN dividers K35 (designed for HV up to 35\,kV) and K65 (max. HV of 65\,kV) were extensively tested and calibrated against the PTB's standard HV divider MT100 and also against each other. The fact that two such devices are available for KATRIN is of a great importance with respect to their mutual cross-checks and overall redundancy. The dividers K35 and K65 were also recently utilised \cite{Kri11} as the reference for calibration of the HV installation at the ISOLDE facility at CERN. To our knowledge the dividers K35 and K65, together with the PTB's divider MT100, represent the three worldwide best dividers in the given HV range.

\subsection{Requirements for an electron standard}
\label{subsec:requirements}

On the contrary to e.g. gamma spectroscopy, where various gamma ray standards \cite{Fir96} are commercially available, the situation in  the electron spectroscopy is much less favourable. To our knowledge there exists no commercial standard of a \emph{solid} electron source with the energy stability required by KATRIN. The consideration of the HV stability constraint in KATRIN expressed in eq.~(\ref{equ:sigma-obergrenze}) leads to the following requirements for a suitable calibration electron source intended for the application at the monitor spectrometer:
\begin{enumerate}
\item The source should provide a well defined discrete electron line, ideally with the energy close to the endpoint of the tritium $\beta$ spectrum.

\item The electron line energy should be stable enough to check the relative stability of the main spectrometer retarding HV set by eq.~(\ref{emu:requirement-relative}) over the time period of at least two months. It should be noted that the foreseen application of the source is the recognition of the HV scale fluctuations on the time scale of minutes to months. On the other hand, the HV scale fluctuations of the order of seconds cannot be recognised with this method. Such short-term fluctuations can be assessed with the help of sophisticated HV ripple probes.

\item  It may seem unnecessary to state such a stability requirement with respect to an electron calibration standard as e.g. the gamma ray standards naturally provide stable gamma lines. Similarly, this holds also for an electron source in gaseous phase. However, in the case of the conversion electron emitted from a solid source the binding energy of the electron is influenced by solid state effects which directly translates into the kinetic energy of the ejected electron. Therefore, the solid source should provide stable chemical environment for the emitted electrons.

\item The electron line should be sharp, i.e. the intrinsic energy spread should be as small as possible. In the case of a natural standard this requirement translates into a very small natural line width $\Gamma$ of the electron line.

\item In case of a natural standard the half-life of the source should be suitable for its use during the KATRIN experimental run of two months. Ideally, the half-life should span more than one run in order to allow for combined analysis of several data taking runs with tritium.

\item For the calibration purposes only the zero-energy-loss electrons are suitable\label{zero-e-loss}, in other words the electron which undergoes an inelastic collision in the source is lost from the useful elastic peak and therefore not usable for the calibration. The count rate of the zero-energy-loss electrons should be high enough in order to provide good statistics in a reasonable measurement time.
\end{enumerate}

\subsection[Implanted \rbkr\ source]{Implanted $^{\textbf{\small 83}}$Rb/$^{\textbf{\small 83m}}$Kr source}

\begin{figure}
 \centering
\includegraphics[width=0.8\textwidth]{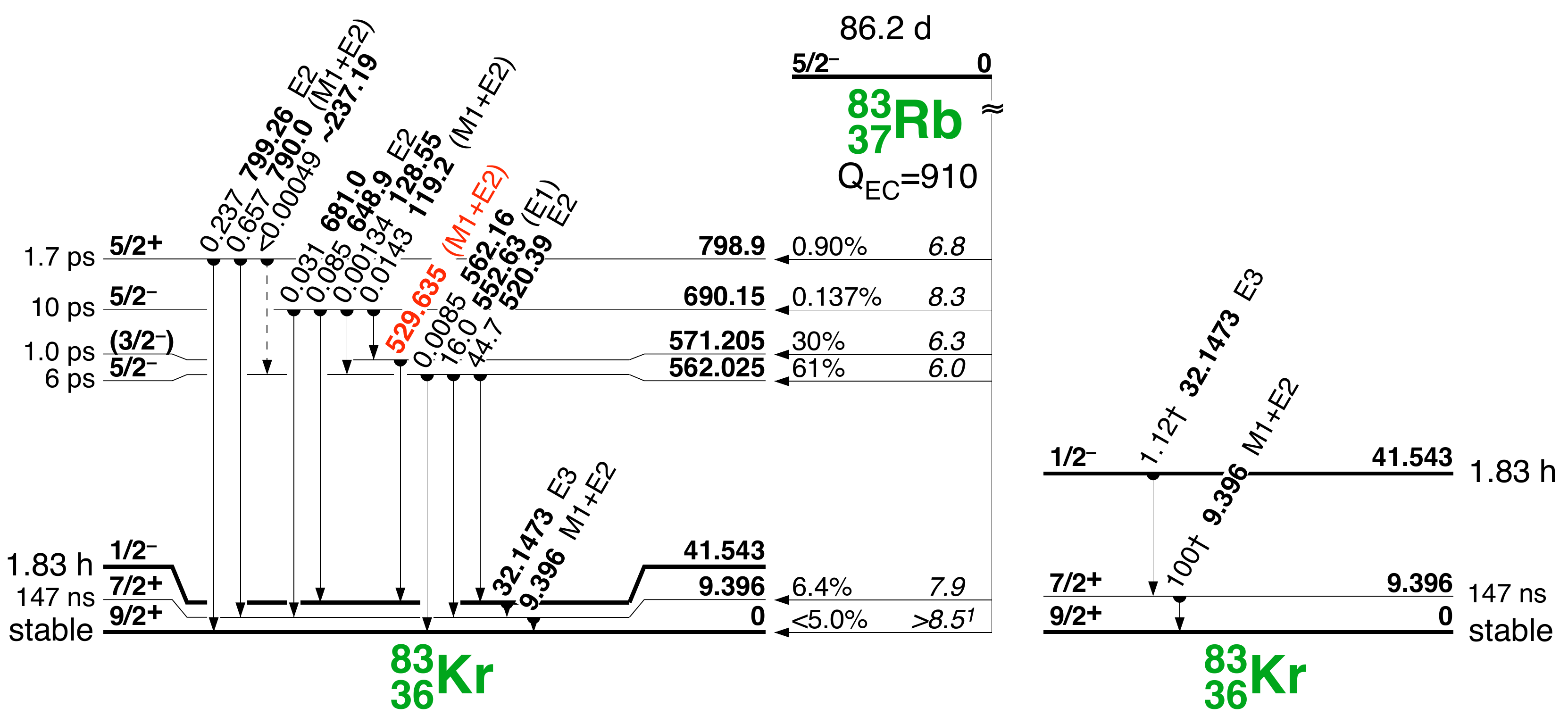}
\caption{The decay scheme of \rb\ to \krs\ (left scheme) and a zoom into the decay of metastable \kr\ to \krs\ (right scheme). The \rb\ decays by a pure electron capture with the branching ratio of 77.9\,\% into the short-lived isomeric state \kr. Detail on the right side shows its 32.2\,keV (intensity of $0.035\,8(45)$\,\%, multipolarity E3) and 9.4\,keV ($5.86(134)$\,\%, M1+E2) gamma transitions. The schemes are taken from \cite{Fir96}.
\label{fig:decay-83rb}}
\end{figure}

An electron source based on the internal conversion of \kr\ was already used in the past for the calibration purposes in Mainz Neutrino Mass Experiment \cite{Pic92} and other neutrino experiments \cite{Rob91,Bel08}. In KATRIN the \kr\ source will be applied in three different physical states \cite{KAT04}: as a gas, as a condensed film of the sub-monolayer thickness and as a solid source. Regardless of the physical state of the given source, the parent isotope \rb\ with the half-life of $T_{1/2}=86.2(1)$\,d \cite{Fir96} will be exploited as \rb\ conveniently serves as a generator (via electron capture) of the short-lived isomer \kr\ with $T_{1/2}=1.83$\,h \cite{Fir96}. The decay scheme of \rb\ is depicted in figure~\ref{fig:decay-83rb}. The conversion electrons produced in electromagnetic transitions from the krypton nuclear levels above the isomeric state are not usable for our purpose due to their high energies. However, the isomeric state decays via a cascade of suitable low energy transitions of 32.2 and 9.4\,keV which possess high intensity of conversion electrons. The first transition (32.2\,keV) with the multipolarity E3 is especially highly converted: The total internal conversion coefficient (ICC) amounts to $\alpha_\mathrm{tot}=2\,010$ \cite{Ros78}. The second transition (9.4\,keV) is practically a pure M1 transition with a small E2 admixture (mixing parameter $\delta=0.013\,0(8)$) and the total ICC amounts to $\alpha_\mathrm{tot}=17$ \cite{Ros78}.

The gamma ray energy of the 9.4 and 32.2\,keV transitions was determined by gamma ray spectroscopy with semiconductor detectors as $E_{32} = 32\,151.7(5)$\,eV \cite{Ven06} and $E_{9.4} = 9\,405.8(3)$\,eV \cite{Sle12}, respectively. The most interesting conversion line of \kr\ regarding the application in KATRIN is the one created by the internal conversion of the 32\,keV transition on the $K$ shell as its kinetic energy is only by 0.8\,keV lower than the tritium endpoint. The precision of the electron binding energy $E_\mathrm{bin}^\mathrm{vac,\;gas}(K) = 14\,327.26(4)$\,eV of this core shell for gaseous krypton was increased in the reevaluation \cite{Dra04}. The kinetic energy of an electron emitted from atomic shell $i$ is roughly given by
\begin{equation}
\label{equ:Kr-gaseous}
E_\mathrm{kin}(i) \approx E_{\gamma} - E_\mathrm{bin}^\mathrm{vac,\;gas}(i)\,,
\end{equation}
where $E_{\gamma}$ is the gamma ray energy and $E_\mathrm{bin}^\mathrm{vac,\;gas}(i)$ is the electron binding energy (related to the vacuum level) of the shell $i$ of a free atom.

The solid type of \kr\ source was introduced in the framework of the KATRIN experiment as a convenient alternative to the proven principle of the condensed \kr\ source (CKrS) \cite{KAT04, Ost08} which is based on earlier work of the Mainz Neutrino Mass Experiment \cite{Pic92}. The concept of such a solid source is based on the fact that the \rb\ generator of \kr\ lies directly in the source itself. Thus, the count rate of the source is driven by the half-life of \rb. The systematic measurements of the energy stability of the \kr\ conversion lines, carried out at the Mainz MAC-E filter spectrometer, were started with \rbkr\ sources prepared by vacuum evaporation of \rb\ onto carbon substrate. In the course of these measurements, which are reported elsewhere \cite{Ven09,Ven10,Zbo11}, it was realised that the sources fulfil the requirements for the time  stability of the conversion line energy but exhibit only moderate retention of \kr\ of about 15\,\%. The retention of \kr\ represents the portion of \kr\ atoms retained in the solid matrix of the source and thus useful for the calibration purposes. In addition, the vacuum-evaporated sources were shown to be very susceptible to surrounding conditions (humidity, vacuum). The idea to implant the \rb\ ions into a suitable substrates was based on the early work \cite{Por71} where the $^{57}$Co ions were implanted at the energy of 500\,eV into cleaved surfaces of natural graphite crystals.

The  implanted \rbkr\ source should be a very promising alternative to the CKrS and even superior in terms of vacuum requirements and ease of use. However, certain change of the electron binding energy is to be expected due to solid state effects\,\footnote{In literature the phenomenon of \emph{core level binding energy shifts of rare gases implanted in noble metals} was studied both experimentally \cite{Cit74,Kim75,Bab92,Bab93} and theoretically \cite{Gad76,Wat76,Wil78,Joh80}. Nonradioactive rare gases (Ne, Ar, Kr and Xe) were implanted into polycrystalline foils of high purity and the XPS method was utilised for measuring the absolute electron binding energies. These were then compared with the binding energies measured for free atoms.}. This can be illustrated in accordance with the approach used in \cite{Cit74}. The kinetic energy of the conversion electron emitted from the atomic shell $i$ of the \kr\ atom placed in a metal host and measured by the spectrometer reads \cite{Sie82}
\begin{equation}
\label{eq:impl}
E_\mathrm{kin}^{\mathrm{impl}}(i)  = E_{\gamma} + E_{\mathrm{rec,\;}\gamma} - E_{\mathrm{rec,\;}e}(i) - E^\mathrm{vac,\;impl}_\mathrm{bin}(i) - \Big(\phi_\mathrm{spec} - \phi_\mathrm{source}\Big)\,,
\end{equation}
where $E^\mathrm{vac,\;impl}_\mathrm{bin}(i)$ denotes the electron binding energy related to the vacuum level. The term $E_{\mathrm{rec,\;}\gamma}$ denotes the energy of the recoil atom after gamma ray emission and is equal to 0.007\,eV and 0.002\,eV for the 32.2\,keV and 9.4\,keV transitions of \kr, respectively. The energy $E_{\mathrm{rec,\;}e}(i)$ of the recoil atom after the emission of the conversion electron from shell $i$ is smaller than 0.22\,eV and 0.12\,eV for the two mentioned transitions in \kr, respectively. The work functions $\phi_{\mathrm{spec}}$, $\phi_{\mathrm{source}}$ of the spectrometer electrode and source, respectively, are important for relating the measured kinetic energy to the retarding potential of the spectrometer. The quantity $E^\mathrm{vac,\;impl}_\mathrm{bin}(i)$ can be expressed in the form
\begin{equation}
\label{eq:impl-2}
E^\mathrm{vac,\;impl}_\mathrm{bin}(i) = E^\mathrm{vac,\;gas}_\mathrm{bin}(i) - \Delta E^{\mathrm{vac}}_{\mathrm{bin}}(i)\,,
\end{equation}
where $\Delta E^{\mathrm{vac}}_{\mathrm{bin}}(i)>0$ represents the solid state correction. This correction related to the Fermi level of the metal host reads
\begin{equation}
\label{eq:impl-3}
\Delta E^{\mathrm{Fermi}}_{\mathrm{bin}}(i)=\Delta E^{\mathrm{vac}}_{\mathrm{bin}}(i)+\phi_\mathrm{source}\,.
\end{equation}
In this case the eq.~(\ref{eq:impl}) can be rewritten into
\begin{equation}
\label{eq:impl-4}
E_\mathrm{kin}^{\mathrm{impl}}(i)  = E_{\gamma} + E_{\mathrm{rec,\;}\gamma} - E_{\mathrm{rec,\;}e}(i) - \Big(E^\mathrm{vac,\;gas}_\mathrm{bin}(i) - \Delta E^{\mathrm{Fermi}}_{\mathrm{bin}}(i)\Big) - \phi_\mathrm{spec}\,.
\end{equation}

The samples investigated in this work were prepared via ion implantation at the ISOLDE facility at CERN. The \rb\ ions were implanted into polycrystalline foils of gold and platinum of high purity. A solid uranium carbide target was used in combination with tantalum surface ion source, as rubidium, being an alkali metal, is easily ionised. In order to maximise the portion of zero-energy-loss electrons a low implantation energy of 30\,keV was chosen, corresponding to the minimal implantation energy available at ISOLDE at that time. In one case, the ion beam was decelerated by retarding HV down to 15\,keV. In all the cases the target foils were kept at room temperature and perpendicular to the ion beam direction. The implantation took place under relatively poor vacuum conditions of $\approx10^{-4}$\,mbar. Prior to the implantation, the foils of 12\,mm diameter were cleaned in reagent-grade methanol by ultrasonics. In addition, the beam focusing was optimised using a non-radioactive Rb beam of high intensity hitting a paper replacing the foil. This was important especially for the implantation at 15\,keV, in which case the foil was placed about 1\,m behind the usual place of beam incident. The option of sweeping the ion beam over the foil was not utilised during our collections. After the completion of collections the foils were stored on air in special holders with plastic caps preventing any wipe-off. It should be emphasised that \emph{no annealing treatment} of any sample was carried out, thus the damage of the lattice by the irradiation was not healed. However, as the collections took place at room temperature it can be assumed that the lattice underwent self-annealing to some extent. It is expected that after its implantation, \rb\ behaves as Rb$^{+}$ and bonds to the environment of the Au or Pt lattice. Some kind of diffusion cannot be excluded, but due to its reactivity \rb\ will probably stay on its original place. Some strong bonding of \kr\ in the solid phase is not expected as within $\approx10^{-10}$\,s after the decay of \rb\ the electron shells are reorganised. However, as the metastable \kr\ exists for $\approx2$\,h before the conversion electron is emitted, the Au or Pt lattice surrounding the \kr\ atom might influence the conversion electrons via weak bonds.

The first \rb\ collection resulted in the 3.3\,MBq source denoted as Pt-30 (30\,keV \rb\ ions hitting the Pt target). The polycrystalline foil of thickness of 40\,$\mu$m used for this collection was not a standard commercial foil. Its purity was determined as 99.7\,\% with the help of the X-ray fluorescence method, the 0.3\,\% impurity being identified as rhodium. For the second series of collections the commercially available foils\,\footnote{The Au foils AU000341/49 and the Pt foils PT000240/55 from the company Goodfellow \cite{Goo09} were used.} of thickness of 25\,$\mu$m, purity of $99.99+$\,\% and temper quality \tqdb{as rolled} were used, with the impurities specified at the level of tens of ppm or less. As the first implanted source Pt-30 exhibited promising results concerning the energy stability of the \kr\ conversion lines, the same implantation energy and foil element was used once more (but the foil was purer in the second case). This led to the sample designated as \pts\ (4.9\,MBq of \rb). However, the two Pt-30 sources were not strictly identical due to the different levels of impurities and implanted doses. In order to investigate other possibilities of the \rb\ implantation the sources Au-30 (3.3\,MBq of \rb\ in gold foil) and Pt-15 (1.9\,MBq of \rb, implantation energy of 15\,keV) were produced as well. The amount of \rb\ present in the sources right after the implantation was deduced from measurements of the \rb\ activity by gamma spectroscopy carried out about 5\,months later.

\subsection[Distribution of the \rb\ atoms in the implanted sources]{Distribution of the $^{\textbf{\small 83}}$Rb atoms in the implanted sources}

\begin{figure}
\centering
\includegraphics[width=\textwidth]{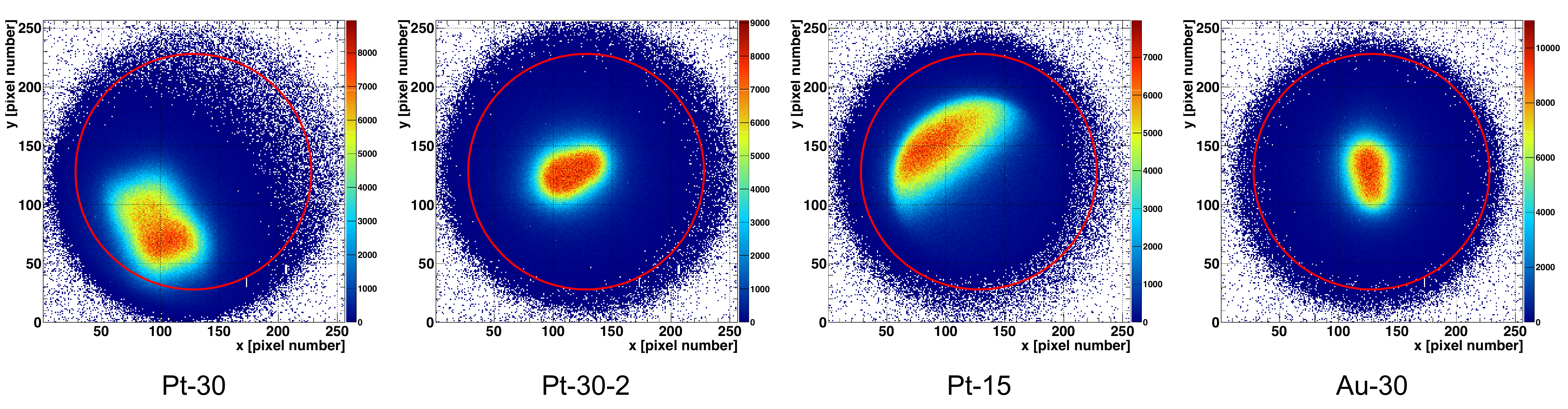}
\caption{The areal distribution of the \rb\ activity in the implanted \rbkr\ sources. The red circles mark the source holders of the inner diameter of 11\,mm. The figure is based on \cite{Sle11}.}
\label{fig:83Rb-xy-profiles}
\end{figure}

The areal distribution of \rb\ atoms over the surface of the four sources was studied \cite{Sle11} with the help of a Timepix detector \cite{Llo07}. The scans of the \rb\ activity are shown in figure~\ref{fig:83Rb-xy-profiles}. It was found that the distribution of \rb\ in the plane of the three sources implanted with a 30\,keV beam can be roughly represented by a two-dimensional Gaussian distribution with a given widths $a,b$ (FHWM). On the other hand, the source Pt-15, produced with the help of the retarding electrode, exhibits a circle-segment-like shape. From the knowledge of the implanted \rb\ activity and the activity spot area the mean implanted dose can be calculated. In all the cases the dose laid in the range of $\approx(1\mbox{\textendash}6)\times10^{14}\,\textrm{ions}/\textrm{cm}^{2}$ (see table~\ref{tab:impl-sources}) at a rate below $2.5\times10^9$\,ions$/$s (corresponding to a current of 400\,pA). From the activity distributions (see figure~\ref{fig:83Rb-xy-profiles}) it can be estimated that the maximal areal dose power did not exceed $220\,\mu\textrm{W}/\textrm{cm}^2$ so that any heat load caused by the beam can be neglected. A dose of the order of $10^{14}\,\textrm{ions}/\textrm{cm}^2$ is generally considered as a limit above which the host lattice damage takes place \cite{Dav80b}. As expected, in the case of the well focused sources \pts\ and Au-30 the implanted dose was the highest. The retention of \kr\ in the implanted sources was found to be $\gtrsim90$\,\% \cite{Sle11}. It may seem surprising that the retention is not 100\,\%, however, it was pointed out in \cite{Dav80b,San06} that a diffusion of the implants to the surface might occur.

The distribution of \rb\ atoms along the depth of the foils was assessed with the help of the SRIM code \cite{SRI11}. Simulations were carried out with the following parameters: The implantation energy was set to 15\,keV (source Pt-15) and 30\,keV (all other sources) and the incident angle (relative to the source foil normal) was set to zero. There is no temperature parameter in the SRIM simulation\,\footnote{The self-annealing effects, occurring at room temperature, are not included in the SRIM calculation \cite{SRI11}.}. Polycrystalline structure of the foils is the default and only option in the SRIM code describing well our situation. Due to poor vacuum conditions surrounding the foils during the implantation, there was certainly a considerable amount of water and hydrocarbons adsorbed on the foil surface. This fact was reflected in the simulations by introducing an additional layer above the metal foil. For the sake of simplicity, this \tqdb{rest gas layer} was set to be a pure carbon layer of the thickness of 3\,nm which was assumed to be a good approximation of the surrounding conditions. Moreover, the sputtering effect, produced by \rb\ ions impinging on the surface, is negligible as the ions possess too high energy and penetrate too deep into the metal foil. This statement is supported by the simulation results: The sputtering yield of 30\,keV Rb ions was found to be in the range of 10\textendash20\,atoms (Pt or Au) per Rb ion which, together with rather low \rb\ ion beam intensities of $\approx2\times10^{9}$\,ions$/$s at ISOLDE, means that less than one monolayer of the rest gas layer was sputtered during the implantation.

\begin{table}
\centering
\caption{Overview of the implanted \rbkr\ sources. $A$ stands for the \rb\ activity right after the source production. The values $a, b$ are the widths (FWHM) of an elliptic shape of the activity spot. Knowing the values $A, a, b$ the mean implanted dose $Q$ was calculated. The simulations carried out with the SRIM code delivered values of the mean projected range $R_\mathrm{p}$ and its standard deviation $\Delta R_\mathrm{p}$. The peak \rb\ concentration $C_\mathrm{p}$ is estimated on the basis of $Q$ and $\Delta R_\mathrm{p}$. Considering the density of lattice atoms, the relative peak atomic concentration $c_\mathrm{p}$ of \rb\ is also calculated. In the last column the measured retention of \kr\ is listed.}
\vspace{7pt}
\begin{tabular}{lrrrrrrrrr}
\hline
source & $A$ & $a$ & $b$ & $Q$ & $R_\mathrm{p}$ & $\Delta R_\mathrm{p}$ & $C_\mathrm{p}$ & $c_\mathrm{p}$ & \kr\ \\
& [MBq] & [mm] & [mm] & [$10^{14}/\mathrm{cm}^{2}$] & [nm] & [nm] & [$10^{20}/\mathrm{cm}^{3}$] & [\%] & ret. [\%] \\
\hline
Pt-30 & 3.3 & 3.8 & 2.6 & 2.3 & 8.9 & 4.2 & 2.2 & 0.33 & 97\\
Pt-30-2 & 4.9 & 3.0 & 1.9 & 5.8 & 8.9 & 4.2 & 5.5 & 0.83 & 94\\
Pt-15 & 1.9 & 5.4 & 2.9 & 0.9 & 6.1 & 2.7 & 1.3 & 0.20 & 88\\
Au-30 & 3.3 & 2.7 & 1.5 & 5.6 & 9.6 & 4.7 & 4.8 & 0.81 & 89\\
\hline
\end{tabular}
\label{tab:impl-sources}
\end{table}

\begin{figure}
\centering
\includegraphics[width=0.6\textwidth]{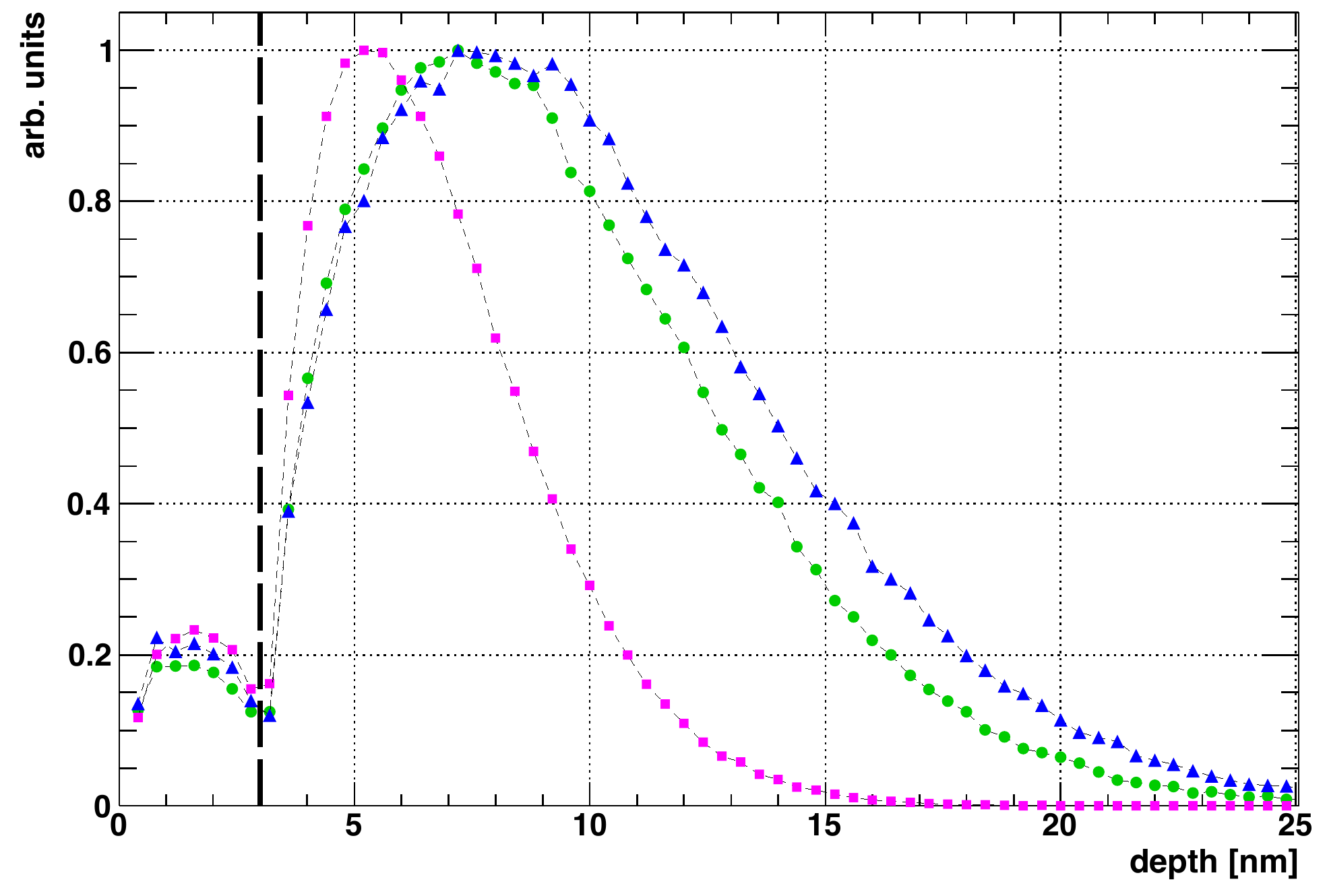}
\caption{The range profiles of Rb ions implanted into gold and platinum, simulated with the help of the SRIM code: Pt-30 (green points), Pt-15 (magenta) and Au-30 (blue). The dashed line at the depth of 3\,nm denotes the boundary between the metal foil and the carbon layer mimicking the rest gas layer which was included in the simulation. The maxima of the distributions were individually scaled to unity.}
\label{fig:imp-profile}
\end{figure}

The implantation ranges, obtained after the simulation of $1.5\times10^{5}$ of ions in each case, are depicted in figure~\ref{fig:imp-profile}. It can be seen that the portion of 5\textendash10\,\% of the incident \rb\ ions gets stopped in the rest gas layer which represents a non-metal environment. As the foils were not cleaned by ion sputtering or any other means after the collections, a certain portion of \rb\ is assumed to remain in this layer. The simulated values of the mean projected range $R_\mathrm{p}$ and the longitudinal straggle $\Delta R_\mathrm{p}$ (standard deviation of the range distribution) are also summarised in table~\ref{tab:impl-sources}. Knowing the values $Q$ and $\Delta R_\mathrm{p}$, via the approximation \cite{Tow76}
\begin{equation}
\label{equ:impl-dose-peak-conc}
Q=\sqrt{2\pi} \cdot C_\mathrm{p} \cdot \Delta R_\mathrm{p}
\end{equation}
one can estimate the peak concentration $C_\mathrm{p}$ of \rb. In all four cases the peak concentration lies in the range of $\approx(1\mbox{\textendash}6)\times10^{20}\,\mathrm{ions}/\mathrm{cm}^{3}$. Considering the density of Pt or Au atoms as $6.62\times10^{22}/\mathrm{cm}^{3}$ or $5.90\times10^{22}/\mathrm{cm}^{3}$, respectively, the estimated \rb\ peak atomic concentrations are found in the range of 0.2\textendash0.8\,\%.
 
\section{High resolution conversion electron spectroscopy with the MAC-E filter}
\label{sec:spectroscopy}

\subsection{Experimental set-up}

\begin{figure}
 \centering
\includegraphics[width=0.8\textwidth]{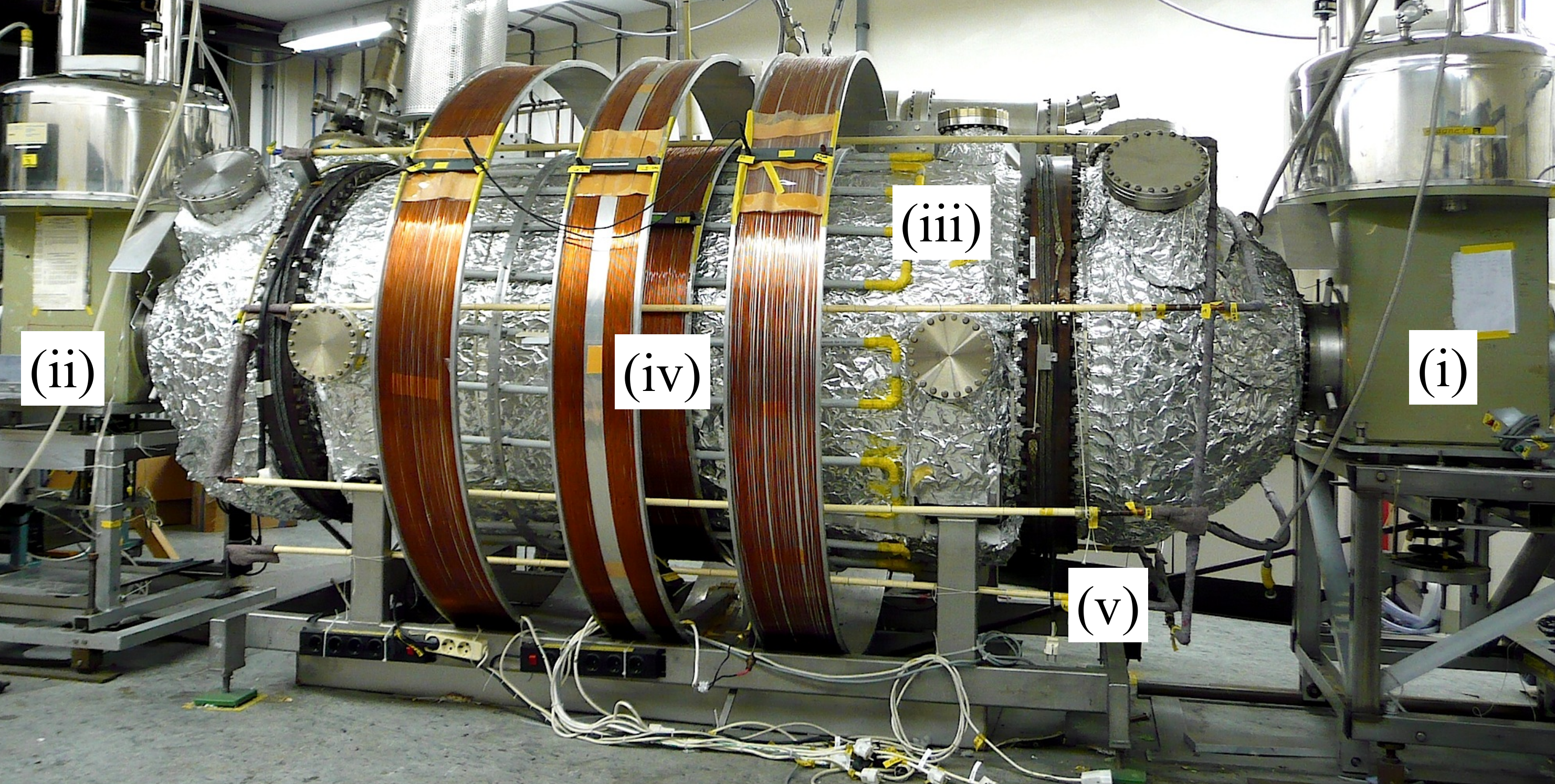}
\caption{The photograph of Mainz MAC-E filter set-up in 2009: (i) solenoid A, (ii) solenoid B, (iii) spectrometer vessel wrapped in thermal insulation, (iv) auxiliary coils for adjusting the spectrometer resolution, (v) auxiliary perpendicular coils for magnetic flux positioning and Earth's magnetic field compensation. The electron sources and the detector were placed in the bores  of the solenoids B and A, respectively.}
\label{fig:mainz}
\end{figure}

The samples of the implanted \rbkr\ sources were investigated with the electron spectrometer of the MAC-E filter type in the Institut f\"{u}r Physik, Universit\"{a}t Mainz. This MAC-E filter, shown in figure~\ref{fig:mainz}, was previously used as the $\beta$ electron spectrometer in the former Mainz Neutrino Mass Experiment \cite{Kra05} until 2001 and later it was upgraded to the high energy resolution\,\footnote{During the Mainz Neutrino Mass Experiment the spectrometer was typically operated at the energy resolution of 4.8\,eV at 18.6\,keV.} of 0.9\,eV at 17.8\,keV. 

The filter consisted of an ultrahigh vacuum tank (length of $\simeq4$\,m, diameter of $\simeq1$\,m), massive conical and cylindrical inner electrodes (stainless steel, DIN 1.4429), two \tqdb{massless} wire electrodes and a magnet system. The vacuum tank and the conical massive electrodes on opposite sides were grounded. The wire electrodes were conductively connected with the massive electrode and a negative analysing voltage of up to $-35$\,kV was applied to them. The work function $\phi_\mathrm{spec}$ of the central analysing electrode was measured as $\phi_\mathrm{spec}=4.4(2)$\,eV \cite{Pic92-NIM}. The magnet system comprised two superconducting solenoids A and B and two sets of auxiliary air coils. The first one, consisting of four coils, was used for adjusting the field $B_\mathrm{min}$, i.e. the energy resolution of the spectrometer. The second set of two windings, perpendicular to each other, allowed to create a magnetic field in any transversal direction. This was used for compensating for misalignments of the source and detector relative to the symmetry axis ($z$ direction) of the spectrometer. Working positions of the source and detector were on the $z$ axis inside the bores of the corresponding superconducting solenoids. The magnetic field in the centre of solenoids amounted to 6.014\,T. The distance between the centres of the solenoids was 4.02\,m. The minimum magnetic field was set to $B_\mathrm{min}=0.3$\,mT in order to achieve the spectrometer energy resolution of $\Delta E = 17.8\,\mathrm{keV}\cdot0.3\,\mathrm{mT}/6\,\mathrm{T}=0.9\,\mathrm{eV}$ at 17.8\,keV, see eq.~(\ref{equ:energy-resolution}).

\begin{figure}
 \centering
\includegraphics[width=0.8\textwidth]{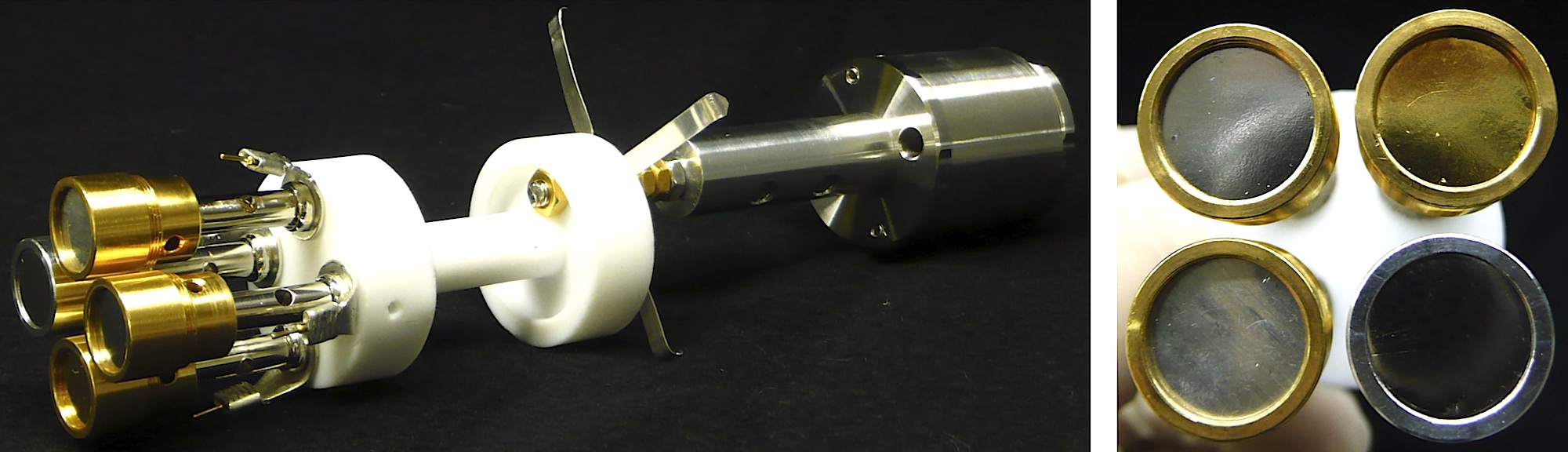}
\caption{The photograph of the source holder allowing to investigate up to four samples without breaking the vacuum. On the right side the detail of the four foils (each in its own sample holder) with the implanted \rbkr\ sources is shown. The gold foil is recognisable in the top right position, the other three foils are the platinum foils. The sample holder in the bottom right position was made from aluminium, whereas the other holders were produced from gold plated copper. The inner and outer diameters of each sample holder were 11 and 14\,mm, respectively.}
\label{fig:4sourcesfull}
\end{figure}

The electron sources were placed in a holder, see figure~\ref{fig:4sourcesfull}. The source holder was mounted on a long rod fixed at the flange. The flange was attached to the $x$-$y$-$z$ table which allowed to centre any of the four sources onto the spectrometer axis. The $x$-$y$-$z$ table was connected via a large diameter bellows to the spectrometer so that the positions of the sources could be changed without breaking the vacuum. The movement in the $z$ direction was motor-driven with the precision of 1\,mm. The source could be positioned in the range of 12.5\textendash70\,cm relative to the centre of the solenoid B. The movement in the $x$ and $y$ directions was done manually with the precision of 0.1\,mm. The elevation amounted to $\pm1$\,cm for both directions. The samples were electrically isolated among each other and also relative to the spectrometer. Each sample was connected to the vacuum feedthrough by a wire which allowed to ground or bias the sample, the maximal applicable voltage was 3\,kV. This way it was possible to separate each electron source energetically (see section~\ref{subsec:typical-meas}). The usual measurement position of the source was chosen as $z=21$\,cm from the centre of the solenoid B, corresponding to $B_\mathrm{s}=1.75$\,T and $\theta_\mathrm{start}^\mathrm{max}=32.7\degree$, cp. eq.~(\ref{equ:pinch-angle}). Larger acceptance angles $\theta_\mathrm{start}^\mathrm{max}$ were avoided in order to limit the energy losses of electrons starting inside the source foil under large angles.

The windowless Si PIN photodiode Hamamatsu S3590-06 with the effective depletion layer of 0.5\,mm thickness \cite{Ham06} was used as the electron detector. Its active area of $9\times9\,\mathrm{mm}^{2}$ was mounted beneath a circular mask of diameter 9\,mm. The detector set-up was mounted on a movable $x$-$y$-$z$ table which allowed to adjust the detector inside the bore of the solenoid A with respect to the spectrometer $z$ axis. The usual position of the detector was $\simeq23$\,cm from the solenoid A centre, i.e. in the magnetic field of 1.06\,T. The detector was placed on a LN$_{2}$ cooled finger together with the preamplifier electronics. The energy resolution was 3\,keV (FWHM) at 17.8\,keV, the noise edge of the photodiode was found to be 6.4\,keV. Thus, the measurements of the low energy conversion lines at 7.5\,keV were still realisable but with losses in efficiency. The detector signals were amplified by a custom-made charge sensitive preamplifier, shaped by a spectroscopy amplifier (Silena 7614, shaping time 0.5\,$\mu\mathrm{s}$) and digitised by a 13-Bit CAMAC ADC (EG\&G ORTEC AD413). Parallel to the detector signals the signals from a pulse generator (EG\&G ORTEC 448, frequency 100\,Hz) were fed to the charge sensitive preamplifier in order to determine the dead time $\tau$ of the system. The frequency stability of the pulser was regularly checked in such a way that the pulser signal was delivered directly to the spectroscopy amplifier instead of the preamplifier signal. In addition, a histogramming memory (EG\&G ORTEC HM413, FERA-bus interface) was utilised, which significantly decreased the dead time. The dead time was determined to be $\tau\simeq40\,\mu$s in the event-by-event readout mode and $\tau\leq10\,\mu$s in the histogramming readout mode. The precise timing of the measurement was based on the $\mu$-scaler (SPEC. 003, frequency 100\,kHz) which was used together with the precision frequency standard (AR 750747, frequency 1\,MHz).

The high precision HV power supply (FuG Elektronik HCP-18-350000, stability 2\,ppm per 8\,h, reproducibility 10\,ppm \cite{FuG06}, voltage ripple 20\,mV peak-to-peak, ppm stability warmup time 2\,h \cite{Thu07}) delivered the HV to the massive and wire electrodes inside the spectrometer. The HV was also fed to the aforementioned high precision HV divider K65, scaling the HV down to 20\,V range. The key property of a HV divider is the long-term stability of its dividing ratio $M(t)$. In general, the long-term dependency of $M(t)$ can be described as exponential \cite{Thu07} which \textemdash\ after sufficient time has passed from the divider production date \textemdash\ can be simplified to a linear function
\begin{equation}
\label{eq:divider-drift}
M(t)=M_0\cdot\Big(1 + \underbrace{\frac{1}{M_0}\frac{\mathrm{d}\,M}{\mathrm{d}\,t}}_{=\,m}\cdot\, t\Big)\,,
\end{equation}
where the long-term drift $m$ is typically of the order of 0.1\,\ppm\ for a high precision device. The long-term stability of the dividing ratio $M_0=1\,818.109\,6(36)$ of the HV divider K65 was extensively studied in the Institut f\"{u}r Kernphysik, Universit\"{a}t M\"unster, over the time period of 13\,months \cite{dipBauer,privBauer}. The dividing ratio drift was determined as $m = -0.13(10)$\,\ppm, i.e. practically compatible with a zero drift.

The low voltage in the 20\,V range was measured by a commercial digital voltmeter (Fluke 8508A, 20\,V DC range, 7\,$^1\!\!/\!_2$-digits resolution mode \cite{Flu08}). The voltmeter was regularly (once per 1\textendash2\,days) calibrated with the help of the 10\,V DC voltage reference (Fluke 732A, stability specified as 3\,ppm$/$year \cite{Flu08} but actually determined as $<0.3$\,ppm$/$year \cite{privBauer}\,\footnote{Such a low drift can be safely neglected in the considerations presented later in this work as the observed drifts of the energies of the conversion lines were higher by at least one order of magnitude.}) which was calibrated at the PTB. This way the offset and the drift of the scale factor (gain) of the voltmeter were monitored and taken into account.

The experience gained at the PTB showed that the main uncertainty in the low voltage measurement comes from cable effects like the resistance and capacity of the coaxial cables used. In addition, the thermoelectric effect gives rise to thermo-voltages (typically of the order of $10^{-5}$\,V$/$K) at the connectors. Although they contribute only on the sub-ppm level they can add up to a significant uncertainty. Therefore we are using special PTFE LEMO cables (twisted pair with shielding) and Cu-Be or Cu-Te connectors while keeping the environmental temperatures stabilised. The electric connections are realised with crimping instead of soldering. As for the HV connections clean contact and insulator surfaces were maintained. At HV above 10\,kV the relative error caused by the thermoelectric effect is in the $10^{-10}$ range and therefore negligible. Instabilities sometimes observed at higher voltages of about $-30$\,kV (used for the measurement of high energy conversion lines of \kr, e.g. \lineLL\ line) can be removed by cleaning the HV insulator. However, erroneous measurements may still occur due to insufficient electrical connections but should be recognised by the redundant monitoring with classical electric techniques and the nuclear standard presented here.

The solid \rbkr\ sources are intended for continuous monitoring of the energy scale stability of the KATRIN spectrometers. This way they should act as an independent stable reference. However, in our test measurements with the sources the HV dividers K35 and K65 were used as a reference. They were regularly calibrated at the PTB and the drifts of their dividing ratios were well known. According to the concept of the continuous monitoring of the energy scale stability in KATRIN, the \rbkr\ source will be biased by $-0.8$\,kV when used at the monitor spectrometer. This way it will be ensured that the \lineK\ conversion line will match the fixed HV of about $-18.6$\,kV applied simultaneously to the main and monitor spectrometers. However, in our case the tested source was grounded and only the HV applied to the spectrometer electrodes was varied. The reason for such a configuration was a minimisation of the HV measurement uncertainties.

\subsection{Typical measurement and fit of a conversion line}
\label{subsec:typical-meas}

The $z$ position relative to the solenoid B centre was common for all the sources. At the beginning of the measurement of each electron source the $x$-$y$ position of the table was found at which the maximal amplitude of the conversion line \lineK\ was observed. Such a procedure ensured that each source was individually centred with respect to the spectrometer $z$ axis and that, in turn, the electrons emitted from each source experienced the same potential dip across the analysing plane. The remaining sources were biased by $+100$\,V, reducing the electron energy by 100\,eV, so that they had a negligible effect on the spectrum measured with a given source. In addition, the signal-to-background ratio was optimised by tuning the currents in the Earth's magnetic field compensation coils which aligned the electron flux inside the spectrometer.

The actual measurement consisted of scanning a chosen electron line of the \kr\ conversion spectrum. The scan was realised by stepping the HV in a suitable region (typically $\pm25$\,V) around the voltage corresponding to the energy of the line. The uniform distribution of the measurement time in this interval was chosen without further optimisation. The stabilisation time of the HV scale due to the initial warmup of the various devices (power supply, divider, digital voltmeter) of the order of $\sim12$\,h was always ensured. Prior to the measurement in each point the HV power supply was set to a given step value $U_\mathrm{step}$ (requested by the control program) and it was let to stabilise for about 5\,s. A typical step size was 0.5\,V. In addition, after large HV steps, e.g. after switching from $-17.8$\,kV (scan of the \lineK\ line) to $-30.5$\,kV (\lineLL\ line), a new measurement was started only about 30\,minutes after the previous scan had finished. Therefore, the complete HV scale was always well stabilised in the ppm precision range.

\begin{figure}
\centering
\includegraphics[width=0.6\textwidth]{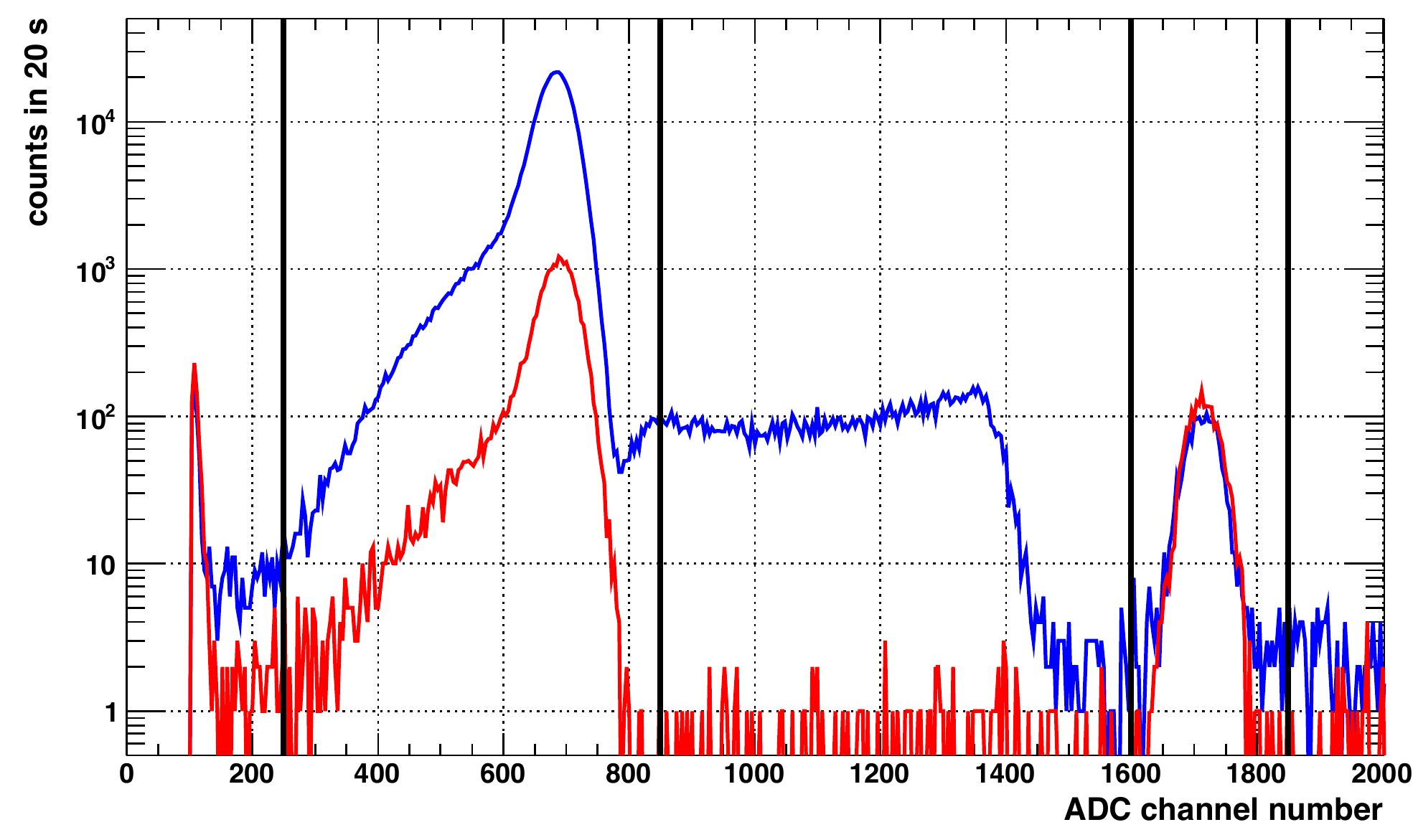}
\caption{Typical raw ADC spectra obtained during the measurement of the \lineLL\ line of the source \pts. The spectrum marked red was recorded in the condition of \tqdb{zero} spectrometer transmission. In this situation the HV applied to the spectrometer electrodes is rejecting all electrons of the \lineLL\ line and only the background \textemdash\ caused by \kr\ conversion lines of higher energies and by secondary electrons emitted from the electrodes \textemdash\ is recorded. The spectrum marked blue was recorded in full transmission, i.e. when all the \lineLL\ electrons are passing the filter HV. The perpendicular lines frame the summation windows in which the electron and pulser (around the channel 1700) peaks are summed in each spectrum. The comparison of the electron peak areas in the depicted spectra results in the signal-to-background ratio of $\simeq18$. As a consequence of the dead time effects due to the high count rate in the case of the full transmission spectrum the pulser peak area is smaller by about 13\,\% in this case. In the full transition spectrum a double electron peak is visible around the channel 1370, stemming from the pile-up effect.}
\label{fig:rawADC}
\end{figure}

For each measurement point at a given $U_\mathrm{step}$ the measurement real time $t_\mathrm{r}$, mean value $U_\mathrm{meas}$ from altogether $N$ low voltage values measured by the voltmeter and a raw ADC spectrum were obtained. Typical ADC spectra are depicted in figure~\ref{fig:rawADC}. The counts in the regions covering the electron peak and the pulser peak were summed up, resulting in peak areas $N_\mathrm{e}$ and $N_\mathrm{p}$, respectively. Due to the periodic nature of the pulser the number of events in this peak does not followed a Poisson distribution but a binomial one. Therefore, in accordance with \cite{Cro05} the variance reads
\begin{equation}
\sigma^2(N_\mathrm{p}) = N_\mathrm{p}\left(1 - \frac{N_\mathrm{p}}{f\,t_\mathrm{r}}\right)\,,
\end{equation}
where $f$ denotes the frequency (nominal value of 100\,Hz) of the aforementioned pulse generator. For the details of the dead time correction see \cite{Zbo11}.

Each HV region was sweep-scanned in both \tqdb{down} (absolute value of negative HV decreasing) and \tqdb{up} (absolute value of negative HV increasing) directions, while both data sets were considered separately as individual integral spectra. One measurement consisted usually of several sweeps. Typically, one sweep took about 0.5\textendash 4\,h, depending on the preset time and the size and number of HV steps. As the radioactive decay of the solid \rbkr\ sources is governed by the half-life of \rb, the time duration of one sweep was negligible with respect to $T_{1/2}(\rb)$ and thus no half-life correction of the counts within the sweeps was necessary.

\begin{figure}
\centering
\includegraphics[width=0.6\textwidth]{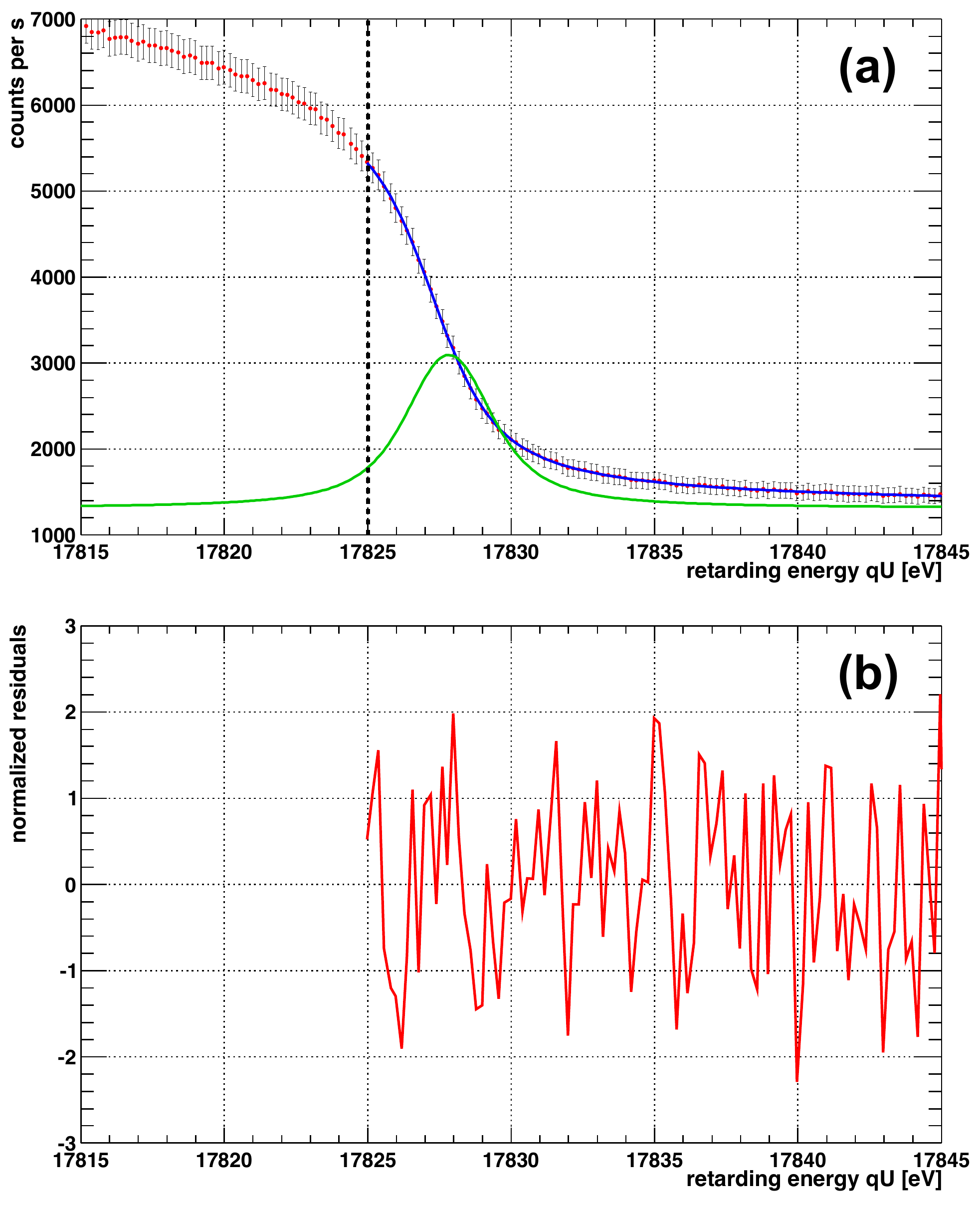}
\caption{Integral spectrum of the \lineK\ line of the source Au-30 and the fit on the range of $[qU_{1}, qU_{2}]=[17\,825, 17\,845]$\,eV. \textbf{(a)} The recorded spectrum corrected for the dead time is shown in red. For the sake of clarity the error bars of the count rate are multiplied by the factor of 10. The blue line denotes the least-squares fit of the convolution of the spectrometer transmission function with a single Voigt profile. The vertical line at 17\,825\,eV marks the lower limit of the fit range. The Lorentz width was kept fixed to $\Gamma=2.7$\,eV, whereas the Gaussian width $\sigma$ was a free parameter. In this case the fit delivered $\sigma=0.77(2)$\,eV. The green line denotes the Voigt profile resulting from the fit (including the constant background $B=1\,313(1)$\cps). The peak centroid was fitted as $E'=17\,827.827(12)$\,eV. \textbf{(b)} The normalised residuals of the fit ($N_\mathrm{dof}=98$, $\chi^{2}_\mathrm{r}=1.10$) shown in \textbf{(a)} exhibit no significant structure.}
\label{fig:579fit-leastsquares}
\end{figure}

A typical integral spectrum \tqdb{dead time corrected count rate in the signal window versus the spectrometer retarding energy} is shown in figure~\ref{fig:579fit-leastsquares}. In general the electron spectrum measured with the MAC-E filter can be expressed as
\begin{equation}
\label{equ:fitcomponents}
g(q U) = R(E_\mathrm{start},q U) \otimes f(E_\mathrm{start}) + B = \big(T \otimes f_\mathrm{trans.\;loss}\big) \otimes \big(f_\mathrm{el.\;loss} \otimes f_\mathrm{shake} \otimes V\big) + B\,,
\end{equation}
where $f(E_\mathrm{start})$ represents the energy distribution of the electrons emerging from the source and $R$ denotes the response of the complete system to monoenergetic electrons. The typical Lorentz distribution $L$ describing an excited atomic state with the natural width $\Gamma$ is smeared by the Gaussian distribution $G$ (broadening due to the Doppler effect and HV fluctuations and possibly due to unresolved chemical shifts), resulting in the Voigt function $V=G \otimes L$. The term $f_\mathrm{shake}$ stands for the shake-up/off phenomena representing the intrinsic mechanisms of the energy losses of electrons\,\footnote{The creation of an inner shell vacancy in the atom \textemdash\ here due to the conversion process \textemdash\ may lead to an additional \emph{simultaneous} excitation or ionisation of outer shell electrons. In the shake-up process a bound electron is excited into an unoccupied bound orbitals. During the shake-off process a bound electron can even be ejected into the continuous spectrum of free electron states. The original electron, ejected from the inner shell (and thus causing the initial inner shell vacancy), is affected by the shake-up/off processes and loses a portion of its energy.}. Further, the loss function $f_\mathrm{el.\;loss}$ describes the effect of the electron inelastic scattering within the source. The inhomogeneities of the electric and magnetic fields\,\footnote{In an ideal {\Mac} both the electrostatic retardation potential $U$ and the minimal magnetic field strength $B_\mathrm{min}$ are invariable along the full extent of the analysing plane. In reality, however, inhomogeneities both in the electric potential and magnetic field will be introduced due to the effects of the finite size of the electrodes and the auxiliary coils as well as local distortions of the field configuration.} can be expressed analytically and included in the transmission function $T$, while $f_\mathrm{trans.\;loss}$ represents the transmission losses which occur typically at high surplus energies as the weak magnetic field in the analysing plane region no longer guides the electrons \cite{Thu07,Pra12}. Finally, $B$ denotes the recorded background rate.

For the purpose of the long-term energy stability measurements of the \kr\ conversion electrons only the zero-energy-loss peak was fitted on the energy range which excluded the shake-up/off peaks, the inelastic scattering of the electrons as well as the effect of spectrometer transmission losses. The fit range was selected also on the basis of Monte Carlo calculations of the elastic and inelastic scattering processes of conversion electrons emitted from the implanted sources \cite{Zbo11}. In this case the spectrum description simplifies to
\begin{equation}
\label{equ:fitcomponents-elastic}
g_\textrm{zero-en.-loss}(q U) = T \otimes V + B\,.
\end{equation}
The function $g_\textrm{zero-en.-loss}(q U)$ was least-squares fitted on the dead time-corrected spectrum with the help of the MINUIT fit routine \cite{Jam75} built in the ROOT environment \cite{ROOT}. For one elastic peak altogether four parameters were used: peak centroid $E'$, line amplitude $A$, Lorentz width $\Gamma$ and Gaussian width $\sigma$. Usually, $\Gamma$ was set fixed to the literature value for a given conversion line, while $\sigma$ was a free parameter.

\subsection{Cross-correlation fit method}

The precise knowledge of the line shape is crucial for a good fit of the conversion line. However, a certain statement about the long-term stability of the electron line energy can be made even without a detailed knowledge of the line shape (as well as the transmission function shape). The method of cross-correlation is often used in signal processing for determining the similarity of two signals. Accordingly, two integral conversion electron spectra were compared with each other: One spectrum, usually measured with high statistics and a fine energy step, was considered as a reference which was then matched to other spectra. The method was utilised in the following manner: Firstly, the reference spectrum, in the case of the \lineK\ line measured with the step of 0.2\,eV, was considered on a range covering the electron elastic peak and background on the high energy side. Secondly, such a template spectrum was least-squares fitted onto the new spectrum (measured with a usual step of 0.5\,eV), while the amplitude $A$, the spectrum position shift $\Delta E'$ and the constant (energy-independent) background $B$ were free parameters. The linear interpolation of the template was necessary in order to obtain its $y$ values in the $x$ points (retarding energies) of the new spectrum. Finally, the goodness of the fit was verified with $\chi^2_\mathrm{r}$ and normalised residuals.

This way a statement about the electron line shift can be easily done as only three free parameters are fitted. However, such a method works flawlessly only if the shape of the conversion line does not change with time. On the contrary to  the vacuum-evaporated \rbkr\ sources where the \kr\ conversion lines could be easily described \cite{Zbo11} with a single Voigt profile, in the case of the implanted sources it turned out to be very difficult to describe the electron spectra reasonably well with a single peak on the full range. However, omitting the low energy side of the peak yielded a good fit, cp. figure~\ref{fig:579fit-leastsquares}. Structures in residuals indicated the presence of a certain asymmetry on the low energy side of the elastic peak. It was possible to fit such an asymmetry with a second Voigt profile with the same widths $\Gamma$ and $\sigma$ as the main peak. The possible origins of the doublet line structure are discussed elsewhere \cite{Zbo11}. Due to certain difficulties of interpreting the low energy side asymmetry and its influence on the fitted centroid of the main peak in the classical many-parameters fit method the cross-correlation fit method was used instead in order to determine the long-term drifts of the conversion lines.

The cross-correlation method was compared with the classical many-parameters fit mentioned above. For the comparison the data of the \lineK\ line ($\approx4\times10^{3}$\cps) of the source Au-30 were taken, measured in well defined experimental conditions spanning 26\,days. A slight discrepancy of $\simeq4\times10^{-7}\,\mathrm{month}^{-1}$ between the \lineK\ line energy drifts obtained by both methods was found. This can be ascribed to the fact that slight changes of the conversion line shape in time like Gaussian broadening, signal-to-background ratio and doublet structure are very likely to occur during the long-term measurements spanning weeks or months. We take this observed difference of 0.4\,\ppm\ as a systematic uncertainty into account when analysing the data with the cross-correlation method.

\section{Experimental results}
\label{sec:results}

\subsection[Drifts of \kr\ conversion lines]{Drifts of $^{\textbf{\small 83m}}$Kr conversion lines}
\label{subsec:drifts}

\begin{figure}
\centering
\includegraphics[width=0.7\textwidth]{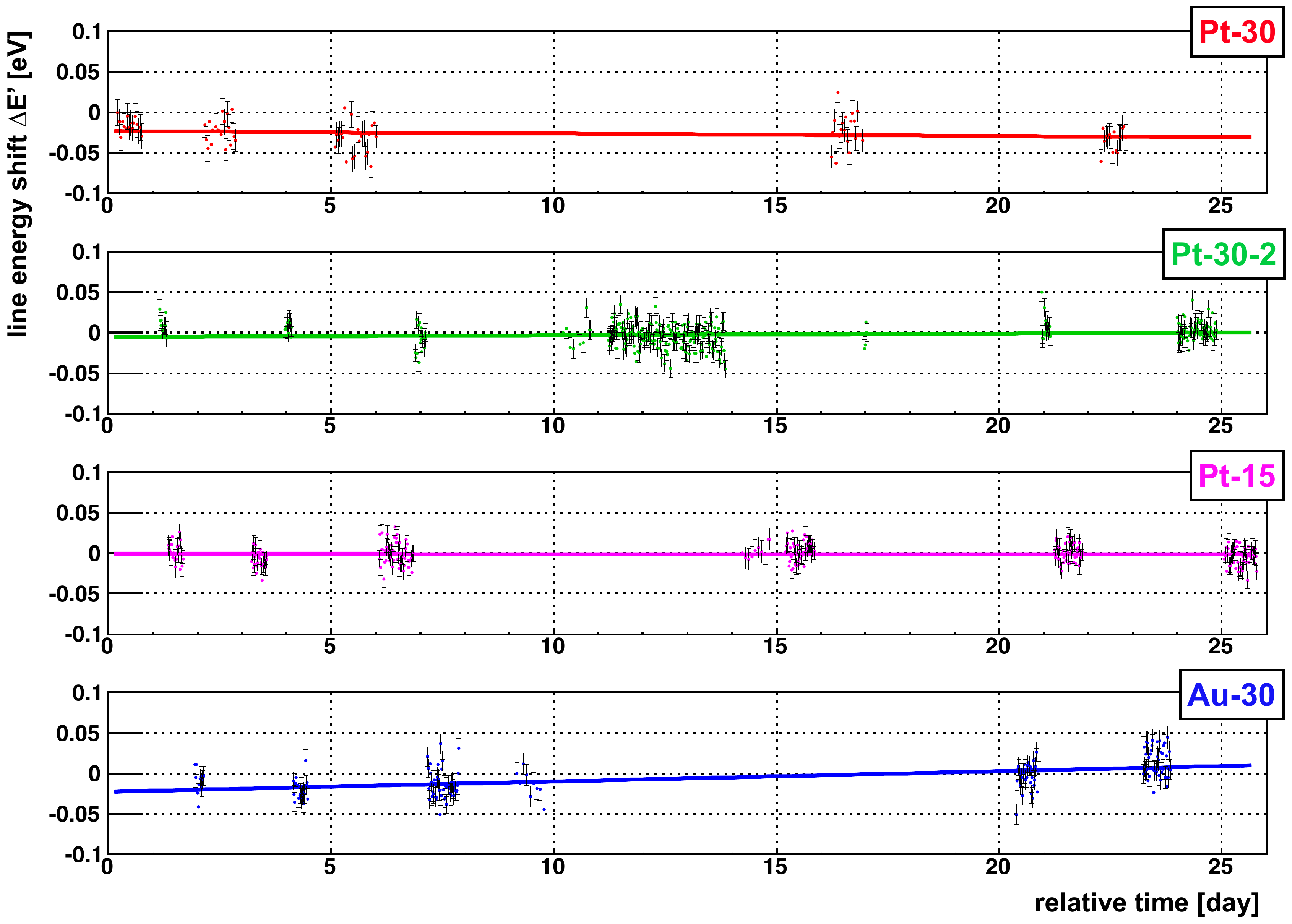}
\caption{The relative positions of the \lineK\ line of the sources Pt-30 (red), \pts\ (green), Pt-15 (magenta) and Au-30 (blue) as a function of time. The linear fit to positions is also indicated, where the Pt-30 data were fitted with $N_\mathrm{dof}=92$, $\chi^{2}_\mathrm{r}=1.27$ and similarly \pts\ ($N_\mathrm{dof}=311$, $\chi^{2}_\mathrm{r}=1.65$), Pt-15 ($N_\mathrm{dof}=248$, $\chi^{2}_\mathrm{r}=1.28$) and Au-30 ($N_\mathrm{dof}=171$, $\chi^{2}_\mathrm{r}=1.87$).  The resulting drifts are summarised in table~\protect\ref{tab:longterm}. The $x$ axis denotes the relative time in days since 27.07.2009. The $y$ axis shows the line energy shift $\Delta E'$ relative to the reference spectrum. The full span of the $y$ axis of 0.2\,eV represents the portion of $\simeq11$\,ppm considering the line energy of 17.8\,keV.}
\label{fig:longterm-stable}
\end{figure}

At the beginning of the measurement all four implanted sources were placed into vacuum simultaneously. About one month prior to the start of the measurements a thorough bake-out of the complete vacuum set-up was accomplished. Moreover, after inserting the sources into vacuum the source section was moderately heated (32\,h at about 80\celsi)\,\footnote{The procedure of moderate heating was intended to reduce the outgassing of the sources and the complete source section. Some influence of heat on the sources cannot be excluded, e.g. the distribution of ions in the host generally changes at elevated temperatures. However, the common practice of removing the host lattice damage via thermal annealing usually requires $2/3\cdot T_\mathrm{melt}$, where $T_\mathrm{melt}$ is the melting point temperature of the host. Therefore, we assume that 80\celsi\ did not significantly affect the distribution of \rb\ ions in the Pt and Au foils.}. The vacuum in the MAC-E filter tank was stable in the $10^{-10}$\,mbar range. The actual measurement consisted of frequent scanning of the most important \kr\ conversion lines \lineK\ (kinetic energy of 17.8\,keV), \lineL\ (7.5\,keV), \lineLL\ (30.5\,keV) and \lineN\ (32.1\,keV)  of each implanted source. The \lineK\ line was measured most often, however, the other lines turned out to be very important for additional systematic measurements. The drifts of the \lineK\ line energy of each source are depicted in figure~\ref{fig:longterm-stable}. The line position shifts were analysed with the help of the cross-correlation method with a reference spectrum for the data of each source. From the beginning of the measurement the \lineK\ line positions of each implanted source exhibited its own unique drift in the \ppm\ range. As each source was actually unique in its properties the differences in the drifts were not surprising. The individual drifts of the sources, corrected for the drifts of the HV divider K65 and the digital voltmeter, are summarised in table~\ref{tab:longterm}. The drifts of the sources Pt-30 and Pt-15 are compatible with zero within $1\sigma$ and $2\sigma$, respectively. The positive drift of the source \pts\ is not compatible with zero which makes this source different from its predecessor Pt-30. Considering the relative drifts, the three sources based on the implantation into Pt foils exhibited the drifts in the sub-ppm range, while the relatively high drift of the source Au-30 amounted to $2.39(19)$\,\ppm. This observation could be attributed to the difference between the noble metals of gold and platinum. The drifts of the other \kr\ lines were studied as well \textemdash\ much less frequently than the \lineK\ line, however, and thus with a lower statistics \textemdash\ and the absolute drift values were found compatible with the drifts of the \lineK\ of the individual sources. 

\begin{figure}
\centering
\includegraphics[width=0.7\textwidth]{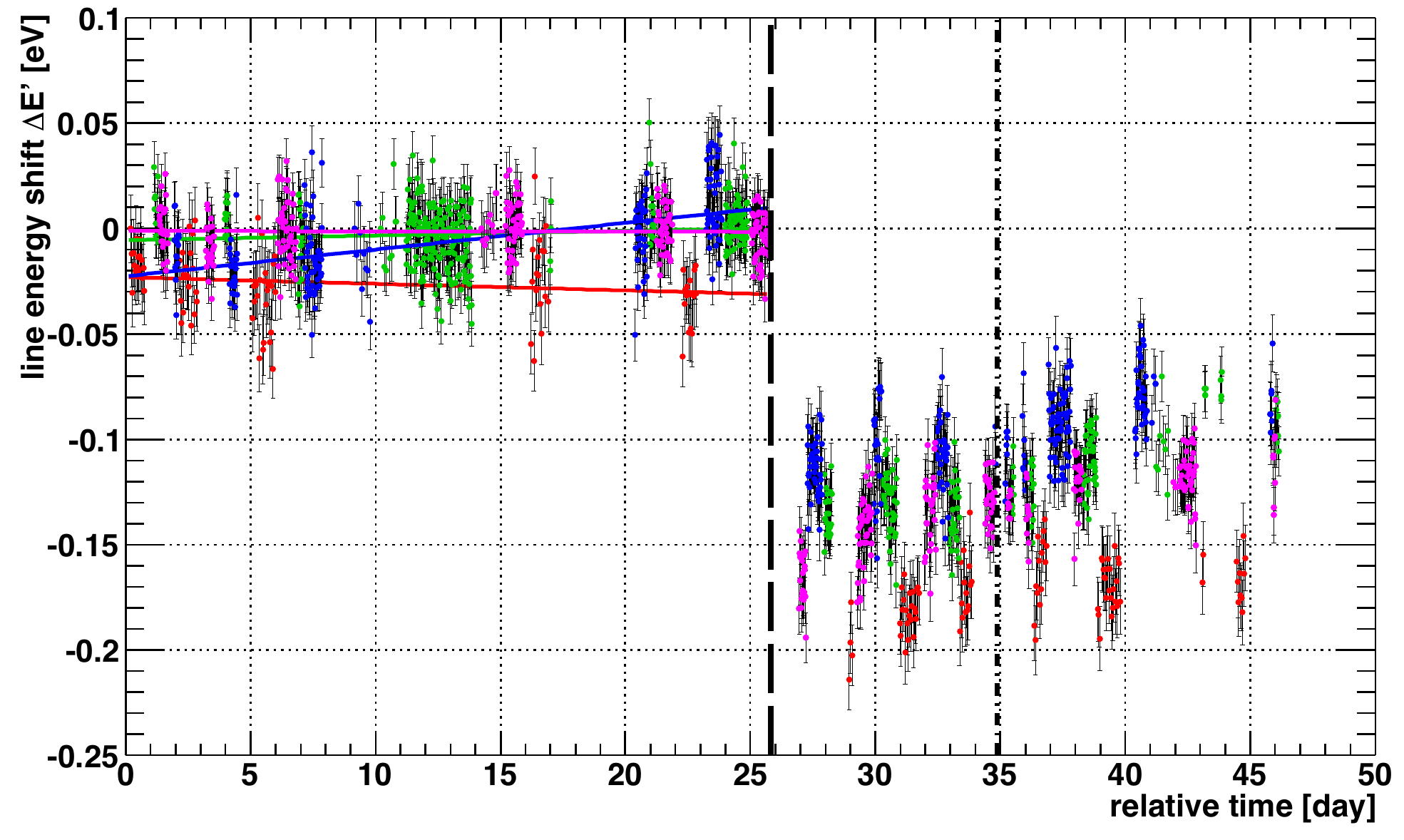}
\caption{The relative positions of the \lineK\ line of the sources Pt-30 (red), \pts\ (green), Pt-15 (magenta) and Au-30 (blue) as a function of time. The $x$ and $y$ axes have the same meaning as in figure~\protect\ref{fig:longterm-stable}. The linear fit to positions on the range of days 0\textendash26 is also indicated. The full span of the $y$ axis is 0.35\,eV ($\simeq20$\,ppm of the \lineK\ line energy). The vertical dashed line on day 26 denotes the vacuum breakdown. The chain-dotted line on day 35 denotes the test of deliberate venting of the source section.}
\label{fig:longterm-with-jump}
\end{figure}

\begin{table}
\centering
\caption{Overview of the relative drifts of the \lineK\ line of the four implanted \rbkr\ sources. The drifts $d_i$ were determined from the analysis of the data collected in days 0--26 only (when the experimental conditions were stable). The drifts $b_i$ stem from the joint analysis of all the \lineK\ line data collected in days 0--46 (including the shift of the work function $\phi_\mathrm{spec}$ due to the vacuum breakdown).}
\vspace{7pt}
\begin{tabular}{|l|rr|rr|}
\hline
source & \multicolumn{2}{c|}{lin. drift $d_i$ (days 0--26)} & \multicolumn{2}{c|}{lin. drift $b_i$ (full data in days 0--46)} \\
& abs.  & rel. & abs. & rel. \\
&  [meV$/$month] & [ppm$/$month] & [meV$/$month] & [ppm$/$month] \\
\hline
Pt-30 & $-5(6)$ & $-0.27(32)$ & $-17(3)$ & $-0.96(17)$\\
Pt-30-2 & $11(3)$ & $0.63(18)$ & $18(3)$ & $1.02(16)$ \\
Pt-15 & $4(2)$ & $0.23(14)$ & $3(3)$ & $0.19(14)$ \\
Au-30 & $43(3)$ & $2.39(19)$ & $47(3)$ & $2.64(16)$ \\
\hline
\end{tabular}
\label{tab:longterm}
\end{table}

Figure~\ref{fig:longterm-with-jump} shows a longer period of data taking with the four sources. The vertical dashed line in figure~\ref{fig:longterm-with-jump} denotes the vacuum breakdown which occurred during the measurement (i.e. all the vacuum sections were affected) on day 26 as a result of the power outage in the whole institute building\,\footnote{Unfortunately, no safety system using some kind of uninterruptible power source was in operation in that moment.}. The vacuum worsened from $10^{-10}$ up to $10^{-4}$\,mbar. The sudden drop of the \lineK\ line positions of all sources is clearly visible in the plot. The weighted average of the \lineK\ line shifts of all the sources amounted to $-0.149(22)$\,eV. In the case of the source \pts\ it was possible to determine the sudden drop of all four most important \kr\ lines (\lineL\ to \lineN) as $-0.157(42)$\,eV (weighted average), independent of the line energy. These observations can be explained as an abrupt \emph{increase} of the work function of the spectrometer electrodes $\phi_\mathrm{spec}$ (common for all the lines of all the sources) upon the residual gas adsorption and a slow recovery of the work function with time during further measurements after the power outage.

In order to verify the hypothesis that the sudden shift of the \lineK\ lines occurred due a change of $\phi_\mathrm{spec}$ several tests of deliberate venting of the source section were carried out. The vertical chain-dotted line in figure~\ref{fig:longterm-with-jump} represents one of such tests. The vacuum in the spectrometer vessel was kept intact (closed by the gate valves) while the source section was vented with air up to atmospheric pressure for about 5\,minutes. Right after the pump-down back to the $10^{-10}$\,mbar range \textemdash\ without any further bake-out of the source section \textemdash\ the stability of the conversion lines was examined. The observed shifts of the \lineK\ line position were compatible with zero: $8(9)$\,meV (source Pt-30), $7(4)$\,meV (\pts), $-3(5)$\,meV (Pt-15) and $-6(12)$\,meV (Au-30). The tests with deliberate venting of the source section showed that the implanted sources are reproducibly stable against abrupt changes of vacuum conditions in their vicinity. The same conclusion was obtained from tests carried out in other time periods (of the investigations of the solid \rbkr\ sources) which were not influenced by any vacuum breakdown.

We clearly see that after day 26 the line position of all four lines comes back to the original position supporting our assumption of a recovery of the work function $\phi_\mathrm{spec}$. In order to disentangle this recovery of the spectrometer electrodes from the drifts of the implanted sources a following approach was chosen: The \lineK\ data of each source were assumed in the form
\begin{equation}
\label{equ:all-drifts}
y_i(t) = \begin{cases}
a_i + b_i\cdot t& \textrm{for $t < t_0$}, \\
a_i + b_i\cdot t + \Delta \phi_\mathrm{spec} \cdot \exp\big[-(t-t_{0})/\tau\big] & \textrm{for $t \geq t_0$}.
\end{cases}
\end{equation}
where $t$ denotes time in days ($t_0 = \textrm{day 26}$), $a_i, b_i$ are the parameters of the linear drift of the \lineK\ line position of the given source $i$, $\Delta \phi_\mathrm{spec}$ represents the shift of the spectrometer work function, common for all the sources, and $\tau$ stands for the \tqdb{lifetime} of the exponential recovery of $\phi_\mathrm{spec}$. The definition of such a function assumes that the drift of $\phi_\mathrm{spec}$ was equal to zero before the vacuum breakdown occurred. In addition, it is assumed that the drift of the given source was not altered by the vacuum breakdown and that the positive drift observed in the time period of days 26\textendash46 is a combination of the \tqdb{original} drift of a given source and the recovery of $\phi_\mathrm{spec}$. The latter one was anticipated to have  an exponential form as a long-term effect of stabilisation of experimental conditions.

\begin{figure}
\centering
\includegraphics[width=0.7\textwidth]{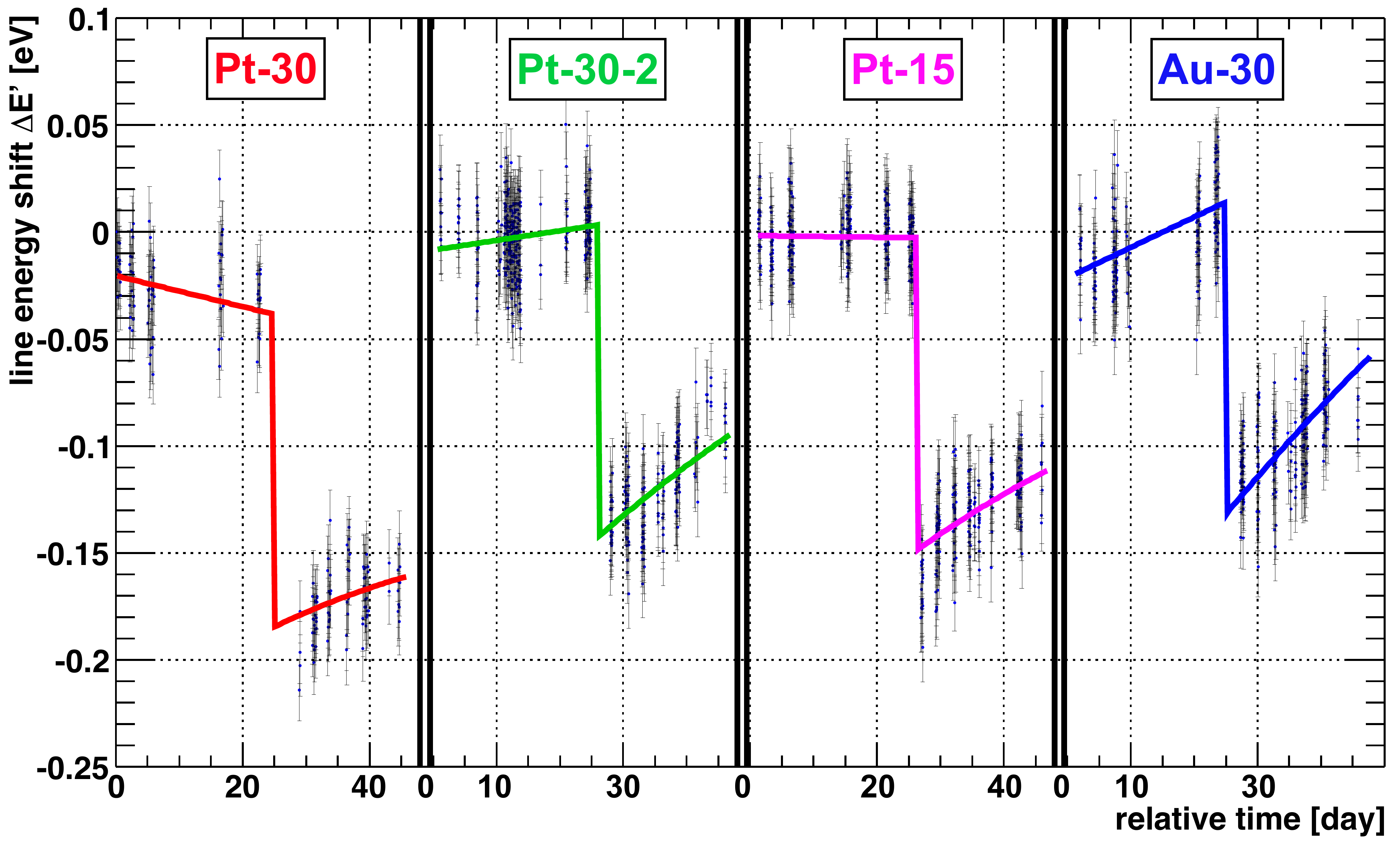}
\caption{The least-squares fit ($N_\mathrm{dof}=1530$, $\chi^{2}_\mathrm{r}=0.96$) of the combination of four functions defined in eq.~(\protect\ref{equ:all-drifts}) to all data of the \lineK\ relative line positions of the four implanted sources. In each source data set the sudden shift $\Delta \phi_\mathrm{spec}$, caused by the vacuum breakdown, is visible as a negative \tqdb{jump}.}
\label{fig:stability-all-at-once}
\end{figure}

A joint fit of the full data set depicted in figure~\ref{fig:longterm-with-jump} was performed: The function from eq.~(\ref{equ:all-drifts}) was used for each source $i$ and altogether 10 free parameters were fitted. Four pairs of the parameters $a_i, b_i$ corresponded to four implanted sources, while $\Delta \phi_\mathrm{spec}$ and $\tau$ were common for all sources. The fit is illustrated in figure~\ref{fig:stability-all-at-once}. The common shift $\Delta \phi_\mathrm{spec}$ was fitted as $-146(2)$\,meV, in a good agreement with the value of $-0.149(22)$\,eV determined from the shifts of individual \lineK\ lines. The lifetime $\tau$ was determined as $68(5)$\,d which illustrates the need for a thorough bake-out of the complete system immediately after such a vacuum breakdown accident. The linear terms $b_i$ [ppm$/$month]  are listed in table~\ref{tab:longterm}. The comparison with the linear drifts determined from the fits of the individual data sets on the range of days 0\textendash26 shows that both approaches deliver comparable results.

\subsection{Systematic effects}
\label{subsec:systematics}

At this point the various systematic effects of the long-term measurements of the conversion electrons energy stability shall be briefly discussed:
\begin{enumerate}

\item the stability of the MAC-E filter energy resolution

Equation~(\ref{equ:energy-resolution}) describes the dependence of the energy resolution on the magnetic fields. The stability of $B_\mathrm{max}$ was monitored by measuring the currents after energising the superconducting solenoids and shortly before de-energising them and this way $B_\mathrm{max}$ was found to be stable at the 0.1\,\% level. The stability of $B_\mathrm{min}$ (typically $B_\mathrm{min}\simeq3$\,G) was estimated by a four days measurement outside of the spectrometer vessel to be better than 0.5\,\%. Both instabilities can be safely  neglected because they do not affect the observed line position at the ppm precision.

\item the stability of the currents in the auxiliary perpendicular coils

The influence of the currents in the auxiliary perpendicular coils on the energy and shape of the conversion line was studied. The parameters of the conversion line (position, amplitude and background) showed a complicated dependence on the currents, however, it became clear that the auxiliary perpendicular coils affect the line shape in the same way as moving the source in the $x$-$y$ plane. Nevertheless, according to the power supplies specifications an instability of the coil currents has a negligible influence on the position of the \lineK\ line. Finally, any malfunction of one of the coils would be immediately observed as a significantly shifted spectrum or as a spectrum with a reduced amplitude-to-background ratio.

\item the position of the source and detector

The influence of the source position on the \lineK\ line was studied with the help of the source Pt-30. The reproducibility of the source positioning in the $x$-$y$ plane was verified. During this the \lineK\ line spectra, recorded after repeated alignment of the source by means of the $x$-$y$-$z$ table, were compared. The set-up was found to be sensitive to the source misalignment: Depending on the direction in the $x$-$y$ plane, the 1\,mm misalignment could cause the shift of the \lineK\ line of $33(3)$\,meV. As the movement precision in the $x$ and $y$ directions amounted to 0.1\,mm, the effect of the source misalignment with respect to the drift of the conversion line energy could be neglected. The $z$ position of the source with respect to the centre of the solenoid B influences the magnetic field $B_\mathrm{s}$ which, in turn, determines the maximal angle $\theta_\mathrm{start}^\mathrm{max}$ of the acceptance cone for the emitted electrons. This $z$ position was fixed during the measurement. The position of the detector with respect to the centre of the solenoid A was chosen in the following way: The $z$ position was chosen as close to the solenoid A centre as possible\,\footnote{Due to mechanical obstacles it was not possible to place the detector in the magnetic fields higher than 1.06\,T.} and the detector was positioned in the $x$-$y$ plane so that the observed background was minimal. Thus, the detector was assumed to lie on the spectrometer $z$ axis. The detector position was kept constant during the measurement.

\item the stability of the spectrometer work function

In the course of the measurement it became obvious that the stability of $\phi_\mathrm{spec}$ is a key factor which could not be monitored in some way at that time. The only means of controlling (minimising) the drift of $\phi_\mathrm{spec}$ was a thorough bake-out procedure performed at the beginning of the measurement. In addition, the simultaneous use (without breaking the vacuum) of the four sources allowed to obtain an indication about a possible drift of $\phi_\mathrm{spec}$ as this term is the same for all of the sources.

\item the drift of the dividing ratio of the HV divider

The drift of $M(t)$ was taken into account in the analysis via the linear term $m$, cp. eq.~(\ref{eq:divider-drift}). The uncertainty of the drifts of the order of 0.1\,\ppm\ seems to be the limit of the systematic uncertainty of the presented method relying on the long-term stability of the HV dividers, however, the dividers proved to be more than sufficiently stable for our purpose.

\item the drift of the scale factor of the digital voltmeter

On the basis of regular calibrations (once per 1\textendash2\,days) the drift of the scale factor $K(t)$ was determined as $k=0.39(9)$\,\ppm. The scatter of the calibration points around the least-squares fitted linear function amounted typically to 0.3\,ppm. The DC 20\,V  range of the voltmeter was calibrated at $-10$\,V with the help of the 10\,V DC reference.
 
\item the stability of the DC 10\,V reference

The stability of the DC 10\,V reference is specified by the producer as 3\,ppm$/$year \cite{Flu08}, however, the drift of the device used in this work was determined as $<0.3$\,ppm$/$year \cite{privBauer}. Such a low drift can be safely neglected in our case.

\item the stability of the scale of the ADC module

Any unrecognised shift of the ADC scale would directly influence the number of recorded electron events which is then used for the construction of the integral electron spectrum. However, the summation windows in the ADC raw spectra were always chosen wide enough (see figure~\ref{fig:rawADC}) so that a small shift of the ADC scale would not matter.

\item the stability of the pulser frequency

The frequency of the research pulser was regularly (weekly) checked and a trend of stabilisation (from $\simeq102.5$\,Hz up to $\simeq105.0$\,Hz) was observed during the time period of 7\,months. In the analysis the mean value was taken into account. The effect of the pulser frequency instability translates into the change of the fitted amplitude and background of a given integral spectrum but it does not effect the fitted line position.
\end{enumerate}

An attempt was made to critically assess all known systematic effects which could possibly affect the presented long-term measurements. Considering the aforementioned effects it seemed appropriate to estimate safely the limit of the experimental systematic uncertainty of the line \lineK\ energy drifts as 0.5\,\ppm. Adding this value quadratically with the discrepancy of  0.4\,\ppm\ stemming from the use of the cross-correlation method the overall systematic uncertainty of the drifts reads 0.64\,\ppm.

\section{Discussion and outlook}
\label{sec:discussion}
\raggedbottom
With the aim of developing a stable monoenergetic electron calibration source we investigated a novel electron source based on \rbkr\ implanted into high purity platinum and gold foils. The conversion lines of \kr\ are suitable for calibration purposes in the KATRIN experiment, namely the kinetic energy of the conversion electrons in the \lineK\ line amounts to 17.8\,keV which is by only 0.8\,keV lower than the tritium $\beta$ spectrum endpoint. At Mainz MAC-E filter we tested the stability of four implanted \rbkr\ electron sources and we showed that they reach the ppm range. The long-term stability of the kinetic energy of the conversion electrons was quantified with the notion of the relative linear drift [ppm$/$month] of the given conversion line of the given source. 

The stability of the electrical devices used in this work, namely the high precision HV divider, the HV power supply and the digital voltmeter, allowed to extract the electron energy drifts from the data. The stability of the spectrometer work function $\phi_\mathrm{spec}$ turned out to be one of the key factors affecting the measurements. Unfortunately, the set-up used in this work did not allow to monitor the stability of $\phi_\mathrm{spec}$ in an independent way. The only means of controlling (minimising) the drift of $\phi_\mathrm{spec}$ was a thorough bake-out procedure performed at the beginning of the measurement. In the time period of 26\,days the drift of $\phi_\mathrm{spec}$ can be safely neglected and the observed drifts of the four implanted sources (see table~\ref{tab:impl-sources}) can be ascribed to the solid state effects in the sources affecting the electron binding energy. The systematic drift uncertainty of 0.64\,\ppm\ was determined on the basis of safe estimates of various effects. All the three Pt-based sources fulfilled the stringent demand of $\pm1.6$\,ppm$/$month on the energy stability $\delta E / E$ set by the KATRIN project. The source Au-30 exhibited non-zero drift and one can speculate about the reason for such a relatively high drift: Due to the well focused ion beam the implanted dose of \rb\ might be too high and serious damage of the lattice could have occurred. However, the same holds for the source \pts\ where the drift was low. This may reflect the differences between the elements of gold and platinum.

The implanted sources were found to be superior to the vacuum-evaporated ones (which are reported in \cite{Ven09}) concerning the resistibility of the conversion line energies to vacuum conditions. Several vacuum breakdowns occurred during the measurement campaign which caused severe temporal worsening of the vacuum from $10^{-10}$\,mbar up to $10^{-4}$\,mbar. In general, these vacuum breakdowns resulted in negative shifts of the conversion line energies measured with the implanted sources. The tests of deliberate venting of the source section separated from the spectrometer vessel made it possible to state that the shift of $\phi_\mathrm{spec}$ is responsible for the observed shift of conversion lines energies. It was found that $\phi_\mathrm{spec}$ was recovering exponentially after the vacuum breakdown. The implanted sources exhibited no shift of the conversion line energy even after exposure to air. This feature was reproducibly tested three times with the precision of 4\,meV. Any vacuum breakdown should certainly be avoided in the future measurements \textemdash\ this holds not only for the monitor spectrometer, but obviously for the whole KATRIN set-up. The fluctuations of the work function will be monitored with the help of dedicated methods on places in the KATRIN set-up where the stability of the work function is important regarding the observable $m^{2}(\nu_{e})$.

In addition, the proper HV connection between the main and monitor spectrometers is of a great importance with respect to the stability of the energy scale in KATRIN: Any unrecognised thermoelectric or cable effect could cause a systematic shift in the monitoring scheme, similar to the shift caused by the work function change. Therefore, special care has to be taken of all cables, feedthroughs and connectors on the HV as well as low voltage side. Recently, different cabling schemes and various locations of the KATRIN HV dividers and HV power supply were investigated with respect to the stability of the \kr\ conversion lines measured at the monitor spectrometer at the KIT. No significant change of the stability was observed even when a HV cable of length of several tens of meters was used \cite{privThuemmler}. Still, the influence of ground loops and RF ripple has to be investigated in the near future when the main spectrometer is running and connected to the monitor spectrometer.

It is worth mentioning that the source Pt-30 exhibited the aforementioned drift, compatible with a zero drift, after almost 9\,months since its production at the ISOLDE facility. In this sense such a solid \rbkr\ source is indeed very well suited for the long-term application in the KATRIN experiment. The four implanted sources showed different drifts probably due to the fact that the sources were produced with different parameters. Therefore no strong statement can be done regarding the reproducibility. This will be the subject of further studies. Still, it is promising that three different sources based on implantation into platinum exhibited the stability compatible with the requirements of KATRIN.

The electron count rate recorded in the \lineK\ line amounted up to $\approx10^{4}$\cps\ in the case of the source \pts. Therefore, the concept of dead time correction utilised in this work was an important part of the data treatment procedure. The amplitudes of the conversion lines generally followed the radioactive decay of \rb\ with the half-life of 86.2\,d, however, significant deviations from this trend were observed. The data analysis revealed that the fitted half-life was actually lower ($T_{1/2}(\textrm{Pt-30})=77.5(4)$\,d, $T_{1/2}(\textrm{\pts})=78.8(3)$\,d, $T_{1/2}(\textrm{Pt-15})=77.2(3)$\,d and $T_{1/2}(\textrm{Au-30})=84.3(3)$\,d) than the literature value. However, here it should be noted only the electrons emitted from the sources without any energy loss were taken into account this way. The lower than expected half-life can point towards the fact that \rb\ diffusion processes take place in the sources. In addition, it cannot be excluded that \kr\ retention of the sources changes in time.

Although very promising results were obtained already with the first samples of the implanted sources, the following details of the production technique could be improved. Stated are also suggestions concerning the future sources which could clarify the open issues. The implanted dose $Q\approx10^{14}\,\textrm{ions}/\textrm{cm}^{2}$ of the current samples shall be lowered by at least one order of magnitude in future. This way the lattice damage induced by the \rb\ ions should be reduced and possibly the line shape shall be less asymmetrical. In order to achieve the \rb\ activity sufficient for long-term application of the source, the option of sweeping the ion beam over the substrate can be utilised at the ISOLDE facility. Thermal annealing of the sources after the \rb\ collection shall be tested (it is also a common procedure to thermally anneal the substrate even prior the implantation). It was realised that according to simulations about 5\textendash10\,\% of \rb\ ions do not reach the metallic state during the implantation, but they end up in the rest gas layer covering the metal foil. Although this fact was not actually proved to be important in the presented measurements, the surface of the metal foil could, in principle, be cleaned by simple means of ion sputtering or plasma etching. Such a procedure would ensure that virtually no \rb\ atoms are present in the non-metallic state. Very promising results were obtained also with the source Pt-15 which, furthermore, profited from reduced portion of inelastically scattered electrons. The option of implantation at low energies ($\approx5$\textendash10\,keV) should be tested as it would increase the number of zero-energy-loss electrons, useful for the energy scale monitoring. As the substrate the polycrystalline platinum foils of high purity seem to be sufficient for our purpose. One could also attempt to implant \rb\ into the HOPG substrate for further reduction of the inelastic scattering of electrons. The use of gold foils does not seem promising with respect to the non-zero shift of the \lineK\ line energy. Finally, the presence of the doublet line structure in the conversion electron spectra remains an issue for the precise description of the line shape \cite{Zbo11}. Further measurements with thermally annealed implanted sources may clarify the origin of the doublet line structure.
\raggedbottom

In summary, the \rbkr\ sources based on implantation into platinum foils provide electron calibration sources with a very high stability in time of the order of 0.1\,\ppm\ over several months. The achievable count rates are high enough to check the stability of the energy scale with 15\,meV precision within several minutes. Therefore such a source is ideally suited as calibration source in the monitor spectrometer of the KATRIN experiment. For the sake of redundancy at least two implanted \rbkr\ sources should be placed in the source section simultaneously and their conversion lines should be measured on a regular basis. We also highly recommend to measure regularly other conversion lines besides the \lineK\ line. We will report about the investigation of the full \kr\ conversion electron spectrum at another place \cite{Zbo12}. This procedure proved to be very useful for checking the stability of the HV scale and the effects of abrupt changes of $\phi_\mathrm{spec}$. However, since the source contains a parent isotope \rb\ with the half-life of 86.2\,d an application in the KATRIN main spectrometer is not foreseen in order to avoid any risk of contamination in case of failure, e.g. by an electric discharge. Its compactness allows the implanted source to be also used in other applications, e.g. for the tests of alignment and transmission properties of the individual components of the KATRIN main beam line.

The presented implanted electron sources can be of interest in other applications where a source of electrons of precise and ultra-stable energy is necessary. The energy resolution should amount to $\Delta E<5$\,eV in such applications in order to separate the zero-energy-loss electrons from the electrons undergoing the inelastic scattering as the latter ones can be influenced by the changes in the source surface.

\acknowledgments

This work was supported by the German Federal Ministry of Education and Research under grant number 05A08PM1 and by the Czech Ministry of Education under grant number P203/12/1896. We wish to thank the members of AG QUANTUM (Institut f\"{u}r Physik, Johannes Gutenberg-Universit\"{a}t Mainz) for their kind hospitality and for giving us the opportunity to carry out these long-term measurements in their laboratory. The visits of A.~Koval\'{i}k to AG QUANTUM were supported by the German Research Foundation under grants number WE-1843/6-1 and 436TSE17/6/06. We are grateful to the ISOLDE collaboration for giving us the opportunity to carry out the implantations of \rb\ (projects I80 and IS500) and to E.~Siesling for his practical help during the collections.

We truly regret that our dear colleague and friend Jochen Bonn deceased during the preparation of this publication. He contributed immensely to the success of the measurements reported here and to the whole KATRIN experiment. We miss him dearly.


\end{document}